\documentclass[a4paper,11pt]{article}
\pdfoutput=1 
\usepackage{verbatim}

\usepackage{jheppub}
\usepackage{amsmath,amsfonts,amssymb,amsthm,graphicx,subfigure,color}
\usepackage[T1]{fontenc} 
\usepackage{enumerate}

\title{\boldmath 	AdS5 solutions from M5-branes on Riemann surface and D6-branes sources}

\author[a,b]{Ibrahima Bah}
\affiliation[a]{Department of Physics and Astronomy, University of Southern California, Los Angeles, CA 90089, USA}
\affiliation[b]{Institut de Physique Th\'eorique, CEA/Saclay, 91191 Gif-sur-Yvette, France}

\emailAdd{bah@usc.edu}

\abstract{
We describe the gravity duals of four-dimensional $\mathcal{N}=1$ superconformal field theories obtained by wrapping $M5$-branes on a punctured Riemann surface.  The internal geometry, normal to the $AdS_5$ factor, generically preserves two $U(1)$s, with generators $(J^+,J^-)$, that are fibered over the Riemann surface.  The metric is governed by a single potential that satisfies a version of the Monge-Amp\`ere equation.  The spectrum of $\mathcal{N}=1$ punctures is given by the set of supersymmetric sources of the potential that are localized on the Riemann surface and lead to regular metrics near a puncture.  We use this system to study a class of punctures where the geometry near the sources corresponds to M-theory description of $D6$-branes.  These carry a natural $(p,q)$ label associated to the circle dual to the killing vector $pJ^++q J^-$ which shrinks near the source.  In the generic case the world volume of the $D6$-branes is $AdS_5 \times S^2$ and they locally preserve $\mathcal{N}=2$ supersymmetry.  When $p=-q$, the shrinking circle is dual to a flavor $U(1)$. The metric in this case is non-degenerate only when there are co-dimension one sources, $M9$-branes, obtained by smearing $M5$-branes that wrap the $AdS_5$ factor and the circle dual the superconformal R-symmetry.  In the IIA limit, they can interpreted as $D8$-branes.  The $D6$-branes are extended along the $AdS_5$ and on cups that end on the co-dimension one branes.  In the special case when the shrinking circle is dual to the R-symmetry, the $D6$-branes are extended along the $AdS_5$ and wrap an auxiliary Riemann surface with an arbitrary genus.   When the Riemann surface is compact with constant curvature, the system is governed by a Monge-Amp\`ere equation. }
\preprint{}

\begin{document} 
\maketitle
\flushbottom

\newpage

\section{Introduction}

$M5$-branes in M-theory have served as a powerful tool for studying and classifying four-dimensional field theories.  In the past, they were used to study the strong coupling regime and IR phases of gauge theories constructed from type IIA branes \cite{Witten:1997sc,Witten:1998jd,Hori:1997ab,Giveon:1997sn}.  In recent times, by studying the IR dynamics of $N$ $M5$-branes wrapped on a punctured Riemann surface, the authors of \cite{Gaiotto:2009we,Gaiotto:2009hg} have provided a classification for $\mathcal{N}=2$ superconformal field theories (SCFTs); these have been dubbed theories of class $\mathcal{S}$.  The typical and generic theory in this classification is strongly coupled and admits no known Lagrangian.  Nevertheless, many of its properties can be inferred from the $M5$-brane realization (see the review \cite{Tachikawa:2013kta} and references therein).  One of the strongest evidence for the existence of the $\mathcal{N}=2$ theories of class $\mathcal{S}$ is the explicit construction of their gravity duals at large $N$ by Gaiotto and Maldacena (GM) in \cite{Gaiotto:2009gz} by using the $\mathcal{N}=2$ Lin, Lunin and Maldacena (LLM) $AdS_5$ systems in M-theory \cite{Lin:2004nb}.  

The primary interest of this paper is to find a way to systematically describe the $\mathcal{N}=1$ SCFTs that arise in the low-energy limit of $M5$-branes wrapped on a punctured Riemann surface in AdS/CFT \cite{Maldacena:1997re}.  This work will have generalized the $\mathcal{N}=2$ GM systems to $\mathcal{N}=1$.  This question has been studied in the context of supersymmetric fields theories in the literature (see \cite{Bah:2011je,Bah:2012dg,Bah:2013aha,Benini:2009mz,Maruyoshi:2009uk,Beem:2012yn,Gadde:2013fma,Xie:2013gma,Xie:2013rsa,Xie:2014yya,McGrane:2014pma,Agarwal:2013uga,Agarwal:2014rua,Bonelli:2013pva,Giacomelli:2014rna} for various approaches and results).  

When $M5$-branes are wrapped on a Riemann surface, the possible configurations can be labelled by the different ways the surface can be embedded in a non-compact Calabi-Yau threefold (CY3).  In particular when the CY3 is a sum of two line bundles over the surface, the resulting theories can be labelled by the degree of the line bundles $(p,q)$.  In the special case when the Riemann surface is compact and has no boundaries, there is a two-parameter family of SCFTs in the low-energy limit of the $M5$-branes.  These field theories and their gravity duals ($B^3W$ solutions) are described in \cite{Bah:2011vv,Bah:2012dg}.  When $p$ or $q$ vanishes, the theories preserve $\mathcal{N}=2$ supersymmetry and its gravity dual is describe by the $\mathcal{N}=2$ Maldacena N\'u\~nez (MN) solution \cite{Maldacena:2000mw}.  

The richness and the wealth of the $\mathcal{N}=2$ class $\mathcal{S}$ theories come from cases when the Riemann surface has punctures.  At these points, boundary conditions are needed for the $M5$-branes; this has the effect of adding global symmetries to the SCFTs.  The addition of the punctures also leads to surfaces with large moduli spaces; in field theory this maps to large conformal manifolds and geometrization of various dualities.  In AdS/CFT these punctures manifest themselves as sources on the Riemann surface in the $\mathcal{N}=2$ MN solutions.  The gravity duals of the class $\mathcal{S}$ theories are obtained by backreacting these sources.  In order to describe them, one needs to first determine the possible choices of sources that preserve the $\mathcal{N}=2$ symmetry and lead to regular metrics.  The backreacted solution is a domain wall that interpolates between a geometry that describes the source and the $\mathcal{N}=2$ MN solution.  GM succeed in this by using the LLM geometry which is governed by an $SU(\infty)$ Toda equation.  

In order to describe the gravity duals of $M5$-branes on punctured Riemann surface, we need to first find the "$\mathcal{N}=1$ version of LLM".  Such system must have an $AdS_5$ factor and the internal manifold must contain a $U(1)^2$ bundle over a two-dimensional space to mirror the basic features of the CY3 in the far UV.  From such structure, we need to find the possible sources on the two-dimensional space that minimally preserve $\mathcal{N}=1$ supersymmetry.  We need to then determine the domain wall solutions that interpolate between the sources and $B^3W$ solutions.  In this paper we address the first step and describe the $\mathcal{N}=1$ system that governs the gravity duals of $M5$-branes on an arbitrary Riemann surface.  We also discuss how to classify the allowed sources\footnote{In \cite{Bah:2013wda}, punctures with $U(1)$ symmetry were studied in $B^3W$ in the probe limit by using $\kappa$-symmetry of $M5$-branes.} and analyse the simplest ones which generalize the $\mathcal{N}=2$ punctures of GM.  We do not, however, describe the solutions that interpolate between the sources and $B^3W$ solutions. This last step may require numerical analysis and we leave it for publications.    

The structure of the paper is as follows.  In section \ref{genM5} we describe the general properties of the field theories from $M5$-branes wrapped on Riemann surfaces.  In section \ref{gravityD} we present the general gravity duals of the field theories and consider various limits related to the LLM system, these are used to describe various punctures.  In section \ref{CCSystems} we analyse the general system in cases where the Riemann surface has constant curvature.  In section \ref{punc} we show how to describe punctures, and  in section \ref{puncD6} we study a class of them that generalize the GM punctures.  Sections \ref{punc} and \ref{puncD6} can be read independently from section \ref{CCSystems}. We end with a summary and outlook in section \ref{conclude}.

\section{Generalities of $M5$-branes on Riemann surface}\label{genM5}

In this section we review some general properties of four dimensional $\mathcal{N}=1$ superconformal field theories obtained by compactifying a stack of $N$ $M5$-branes on a Riemann surface with genus $g$ and $n$ punctures.  The surface is given by a complex co-dimension two curve, $\mathcal{C}_{g,n}$, of a non-compact Calabi-Yau threefold, $CY_3$.  Equivalent, we can view the four dimensional field theory as the low-energy limit of the six dimensional $A_{N-1}$ $(2,0)$ field theory on $\mathcal{C}_{g,n}$.

\subsection{Twist and symmetries} \label{classS}

Supersymmetry is generically broken when supersymmetric field theories are taken over curved manifolds.  For the case of interest, we can preserve some supersymmetry by a partial topological twist \cite{Witten:1988ze,Bershadsky:1995qy}.  We look for a constant spinor by turning on a background gauge field valued in the R-symmetry of the field theory and tune it to cancel the background curvature.  In other words we pick a gauge field, $A$, to solve the killing spinor equation
\begin{equation}
\left(\partial_\mu + \frac{1}{4} \omega^{ab}_{\;\;\; \mu} \Gamma_{ab} + A \right) \varepsilon = \partial_\mu \varepsilon =0.
\end{equation} where $\omega^{ab}_{\; \mu}$ is the spin connection of the manifold and $\varepsilon$ is the desired spinor.  

The possible choices of twists we can perform is determined by the holonomy group acting on the spinor, where $\omega^{ab}_{\;\;\; \mu} \Gamma_{ab}$ takes value, and the different ways we can embed it in the R-symmetry.  For the case of interest, we have a two dimensional curved manifold and therefore a $U_h(1)$ holonomy.  The $(2,0)$ theory has an $SO(5)$ R-symmetry; we thus consider its abelian subgroup
\begin{equation}
U_+(1) \times U_-(1) \subset SU_+(2)\times SU_-(2) \subset SO(5)
\end{equation} and identify the holonomy group with a linear combination as
\begin{equation}
U_h(1) = \frac{p'}{p'+q'} U_+(1) + \frac{q'}{p'+q'} U_-(1).  
\end{equation} The topological twist fixes the combination of the parameters\footnote{We reserve $(p,q)$ as the twist parameters on a compact Riemann surface where $p+q=2(g-1)$.} as
\begin{equation}
p'+q' = 2(g-1) +n. \label{twist}
\end{equation}  This is the condition that the flux of the background gauge field must cancel the curvature of the Riemann surface which is given by its Euler characteristic $\chi = 2(1-g) -n$.  The twist generically preserves four supercharges in four dimensions and therefore leads to $\mathcal{N}=1$ field theories.  The bosonic symmetries of the six dimensional theory is broken down as
\begin{equation}
SO(1,5) \times SO(5) \; \to \; SO(1,3) \times U_+(1) \times U_-(1).
\end{equation}  The generators of the $U_\pm(1)$ are denoted as $J^\pm$.  The field theory admits an R-symmetry, $R_0$, and  a flavor symmetry, $\mathcal{F}$, defined as
\begin{equation}
R_0 = \frac{1}{2} \left(J^+ + J^-\right), \qquad \mathcal{F} = \frac{1}{2} \left(J^+ -J^-\right).  
\end{equation} When the four dimensional field theory is superconformal, we can write the superconformal R-symmetry, $R_{\mathcal{N}=1}$, as 
\begin{equation}
R_{\mathcal{N}=1} = R_0 + \epsilon \mathcal{F}= a_+ J^+ + a_- J^-, \qquad a_\pm = \frac{1}{2} \left(1\pm \epsilon \right).  
\end{equation}  The $\epsilon$ parameter is fixed by a-maximization \cite{Intriligator:2003jj}.

\subsubsection*{Systems with enhanced Symmetry}

\paragraph*{$\mathcal{N}=4$} In the special case when $p'=q'=0$ the Riemann surface is a torus.  In this case, the four dimensional theory is nothing but $\mathcal{N}=4$ Super Yang-Mills (SYM) theory with $SU(N)$ gauge group.  This compactification provides a natural origin of the $S$-duality in $\mathcal{N}=4$ SYM as the modular group of the torus.  The gauge coupling is given by the modular parameter of the torus.

\paragraph*{$\mathcal{N}=2$} When $p'$ or $q'$ vanish, the four dimensional theory preserves eight super charges. These theories have been studied in \cite{Witten:1997sc,Gaiotto:2009we,Gaiotto:2009hg}.  When $p'=0$ ($q'=0$) the $U_+(1)$ ($U_-(1)$) is enhanced to $SU_+(2)$ ($SU_-(2)$).  The $SU(2)\times U(1)$ R-symmetry of the $\mathcal{N}=2$ theory in the two cases can be identified as
\begin{align}
&p' =0: \qquad SU_+(2) \times U_-(1) \\
&q' =0: \qquad U_+(1) \times SU_-(1).  
\end{align}
When the $\mathcal{N}=2$ theory is described in $\mathcal{N}=1$ language, we need to identify an $\mathcal{N}=1$ slice of the $\mathcal{N}=2$ superconformal algebra.  The R-symmetry on the slice can be written as
\begin{equation}
R_{\mathcal{N}=1} = \frac{1}{3} R_{\mathcal{N}=2} + \frac{4}{3} I_3
\end{equation} where $R_{\mathcal{N}=2}$ an $I_3$ are respectively the generators the $U(1)$ and the cartan $U(1)$ of the $SU(2)$ in the $\mathcal{N}=2$ R-symmetry.  This allows us to identify $\epsilon$ for the $\mathcal{N}=2$ theories:
\begin{align}
&p' =0: \quad J^+ = 2 I_3, \quad J^- = R_{\mathcal{N}=2}, \quad \epsilon = \frac{1}{3} \label{n2p'}\\ 
&q' =0: \quad J^- = 2 I_3, \quad J^+ = R_{\mathcal{N}=2}, \quad \epsilon = -\frac{1}{3}  \label{n2q'}
\end{align}

\paragraph*{$\mathcal{N}=1$ with $SU_F(2)$} In the special case when $p'=q'$, $U_h(1)$ is identified with $U_+(1) + U_-(1)$.  This twist preserves the diagonal $SU_F(2)$ of $SU_+(2) \times SU_-(2)$.  For these cases, The flavor symmetry $\mathcal{F}$ is enhanced to $SU_F(2)$.  This class of theories with $n=0$ were studied in \cite{Benini:2009mz}.

\section{Gravity duals}\label{gravityD}

In this section, we describe the gravity duals of the class of superconformal field theories that describe the low energy dynamics of $M5$-branes on a Riemann surface.  First we argue for the general ansatz for such system and then describe the metric.  In appendix \ref{metricsystem} we provide a derivation of the metric following from \cite{Bah:2013qya}.  

\subsection{Ansatz}

In M-theory, we decompose the spacetime as
\begin{equation}
M^{1,10} \; \to \; \mathbb{R}^{1,3}\times \mathbb{R} \times CY_3. \label{CY3system}
\end{equation}   The stack of $N$ $M5$-branes is extend along $\mathbb{R}^{1,3}$ and along a curve $\mathcal{C}_{g,n}$ in $CY_3$.  In general, the $CY_3$, near the brane region, is a $U(2)$ bundle over $\mathcal{C}_{g,n}$ whose determinant line bundle is fixed to the canonical bundle of the surface.  In this paper we restrict to the case where the structure group of the bundle is $U(1)^2$ ( See section 2 of \cite{Bah:2012dg} for more details).  In this case the $CY_3$ is a sum of two line bundles over the curve with degrees  $p'$ and $q'$ respectively.  The local geometry is 
\begin{equation}
\begin{array}{ccc}
\mathbb{C}^2 & \hookrightarrow & \mathcal{L}_{p'} \oplus \mathcal{L}_{q'} \\ & & \downarrow \\ & & \mathcal{C}_{g,n}.
\end{array}
\end{equation} The vanishing of the first chern class of the $CY_3$ implies \eqref{twist}.  The symmetries $U_+(1) \times U_-(1)$ are the phases of the two line bundles.   

Our primary interest is to understand when these configurations of $M5$-branes flow to superconformal field theories in the IR.  In the large $N$ limit, this question can be studied in AdS/CFT by classifying the set of $AdS_5$ solutions that can exist in the near horizon limit of geometries with these brane sources.  The problem of obtaining the fully backreacked solution in M-theory is hard.  Since we are only interested in cases when the near horizon geometry contain $AdS_5$ factors, we look for those solutions directly.  The ansatz for the eleven dimensional geometry is then a warped product of $AdS_5$ and a six-dimensional compact space, $M_6$.  The main challenge is to then classify all $M_6$ that are of the form
\begin{equation}
\begin{array}{ccc}
M_4 & \hookrightarrow & M_6 \\ & & \downarrow \\ & & \Sigma_{g,n}.
\end{array}
\end{equation} where $\Sigma_{g,n}$ is IR limit of the curve $\mathcal{C}_{g,n}$.  The four manifold, $M_4$, admits at least a $U(1)^2$ isometry group corresponding to the phases of the line bundles, $U_+(1) \times U_-(1)$.  The circles dual to the generators of $U_+(1) \times U_-(1)$ are non-trivially fibered over the Riemann surface, $\Sigma_{g,n}$.  The eleven dimensional metric must be of the form
\begin{equation}
M^{1,10} \quad \to \quad AdS_5 \times_w\left( \Sigma_{g,n} \times S_+^1 \times S_-^1 \times [t^+] \times [t^-] \right) \label{ansatz}
\end{equation} where $S_\pm^1$ are the circles dual to $U_\pm(1)$.  The last two directions with coordinates $t^\pm$ do not, generically, correspond to any isometries.  The $AdS_5$ radius and the two interval directions $t^\pm$ are combinations of the line, $\mathbb{R}$ in \ref{CY3system}, and the radii of the fibers in the line bundles.  In general the metric on $M_6$ will depend on the coordinates on the Riemann surface and on $t^\pm$, therefore one expects a system of equations with four variables when we reduce the supergravity equations.  While this is true, we will find the system to be tractable.  

The metric for the most general supersymmetric $AdS_5$ solution in M-theory \cite{Gauntlett:2004zh} has the form
\begin{equation}
ds^2_{11} = H^{-1/3} \left[ ds^2_{AdS_5} + H \left( ds^2_4 + \frac{dy^2}{1-4y^2 H} + \frac{1}{9} \frac{1-4y^2 H}{H} \left(d\psi + \rho\right)^2\right) \right].
\end{equation} The warp factor, $H$, and the metric depends on $y$ and on the coordinates on $ds^2_4$; at fixed $y$ the four-dimensional space is K\"{a}hler.  The supergravity equations for the system can be written in terms of the K\"ahler and holomorphic two-forms on $ds^2_4$.  The $\psi$ direction parametrizes a circle fibered over the K\"ahler space; its connection, $\rho$, is completely fixed by the base four manifold.  The $U(1)$ generated by $\partial_\psi$ is dual to the superconformal R-symmetry.    

For the problem of interest, we make the ansatz
\begin{equation}
ds^2_{4} = e^{2A}\left(dx_1^2 + dx_2^2\right) + e^{2B} \left[\left(dz+V^R \right)^2 + e^{2C} \left(d\phi+ V^I\right)^2 \right].  
\end{equation} The $x_i$ coordinates parametrize the Riemann surface directions.  In addition to the R-symmetry circle $\psi$, we also demand a circle on the base manifold in order to obtain the desired $U(1)^2$ isometry.  The one forms $V^{I/R}$ have legs on the Riemann surface.  All metric functions depend on $x_i$, $z$ and $y$.  The reduction of the BPS equations on the ansatz is worked out in \cite{Bah:2013qya}.  In appendix \ref{metricsystem} we summarize the results.   

We can identify the isometry $\partial_\phi$ with the flavor symmetry in the dual theory.  From this we obtain
\begin{equation}
\partial_\phi = \frac{1}{2} \left(\partial_{\phi_+} - \partial_{\phi_-}\right), \qquad \partial_\psi = a_+ \partial_{\phi_+} + a_- \partial_{\phi_-}
\end{equation} where 
\begin{equation}
J^\pm \qquad \mbox{is dual to } \qquad \partial_{\phi_\pm}.  
\end{equation}  By using this identification, we can make field redefinitions and coordinate transformations on the system in \cite{Bah:2013qya} to obtain a more useful description of the gravity duals to the class of SCFTs describe in section \ref{classS}.  The reduction of the system in \cite{Bah:2013qya} is described in appendix \ref{metricsystem}. 

\subsection{Gravity dual to class $S$ SCFTs} \label{Gravitydual}

The gravity dual to $\mathcal{N}=1$ class $S$ theories can be written as
\begin{equation}
ds^2 = L^{4/3} H^{-1/3} \left[ ds^2_{AdS_5} + \frac{1}{3} H \left( e^{2A} \left(dx_1^2 + dx_2^2\right) +4 h^{ij}\eta_i \eta_j + g_{ij} d t^i  dt^j \right) \right] \label{metricgen}
\end{equation} with
\begin{equation}
g_{ij} = - \partial_i \partial_j D_0 , \qquad h_{ij} = - \partial_i \partial_j \left(D_0 + \frac{3}{2} s \left(\log s -1\right)  \right).
\end{equation} and 
\begin{equation}
e^{2A} = s^{5/2} \det(g_{ij}) e^{(\partial_++\partial_-)D_0}, \qquad H = \frac{1}{8s} \frac{\det(h_{ij})}{\det(g_{ij})}, \qquad s =   a_+ t^+ +  a_- t^-.  
\end{equation} The metric along the circles, $h^{ij}$, is obtained from the relation $h^{ij}h_{jk}= \delta^i_k$.  The one-forms dual to the isometry generators $\frac{\partial}{\partial \phi_\pm}$ are given as
\begin{equation}
\eta_{\pm} = d \phi_{\pm} + \frac{1}{2} * d \partial_\pm D_0
\end{equation} where the $*$ and $d$ are taken along the $(x_1,x_2)$ directions.  Our convention for the hodge star is $*dx_1 = -dx_2$ and $*dx_2 = dx_1$.

The length scale of the system is the $AdS$ radius $L$, and it appears as an overall factor, it is fixed as $L=1$ for the rest of the discussion.  The four-form flux is completely determined in terms of the metric functions \cite{Gauntlett:2004zh}; solving for $D_0$ will also determine the flux.  In this paper, we focus on solutions and will not need the flux; it will be necessary for central charge computations.  

The system is governed by a single function $D_0$ that satisfies a variation to the Monge-Amp\`ere equation\footnote{The fact that the general systems in \cite{Gauntlett:2004zh} has underlying Monge-Amp\`ere equations was first described in \cite{Lunin:2008tf}.}
\begin{align}
\left(\partial_{x_1}^2 + \partial_{x_2}^2 \right) D_0 &=  s^{5/2}\left(\partial_+^2 D_0 \partial_-^2 D_0 - \left(\partial_+ \partial_- D_0\right)^2 \right)e^{ \left( \partial_+ +\partial_- \right) D_0 }.  \label{MAeq} 
\end{align}  Our central goal is to understand the solution space of this equation that lead to regular metric.  

\subsection*{ S-duality}

The system is invariant under the transformation
\begin{equation}
\left(\begin{array}{c} \eta_+ \\ \eta_- \end{array} \right) = A \left(\begin{array}{c} \eta'_+ \\ \eta'_- \end{array} \right), \qquad \left(\begin{array}{c} t^+ \\ t^- \end{array} \right) = A^{-1} \left(\begin{array}{c} t'^+ \\ t'^- \end{array} \right) \label{symmetry}
\end{equation} where $A$ is given as
\begin{equation}
A = \frac{1}{2}\left(\begin{array}{cc} 1+r_1 & 1+r_2 \\ 1-r_1 & 1-r_2 \end{array} \right), \qquad \mbox{with} \qquad r_1-r_2 \neq 0.  
\end{equation}  The elements in the columns must add up to one in order to preserve the exponential $\exp\left((\partial_+ + \partial_-) D_0\right)$ in the conformal factor $e^{2A}$.  In order to preserve the $s$ function, $a_\pm$ transform as
\begin{equation}
\left(\begin{array}{c} a_+ \\ a_- \end{array} \right) = A \left(\begin{array}{c} a'_+ \\ a'_- \end{array} \right).
\end{equation} This transformation preserves the condition $a_+ + a_-=1$. However the parameter $\epsilon$ transforms as
\begin{equation}
\epsilon = \frac{r_1+r_2}{2} + \frac{r_1-r_2}{2} \epsilon'. 
\end{equation}  In the field theory side, this transformation corresponds to rotating the generators $(J^+,J^-)$ while preserving the overall combination $J^+ + J^-$.  This invariance in the metric corresponds to an ambiguity on the definition of $J^\pm$.  The only physical input is the topological twist which fixes an overall $U(1)$ in the $SO(5)$ R-symmetry of the six-dimensional $(2,0)$ SCFT.

\subsection{$\mathcal{N}=2$ Limit and LLM}

The system above must reduce to Lin Lunin and Maldacena system (LLM) \cite{Lin:2004nb} in order to describe the $\mathcal{N}=2$ theories.  In the discussion of enhanced supersymmetry in section \ref{classS}, the $\mathcal{N}=2$ limit corresponds to trivializing one of the line bundles.  In Gravity, this corresponds to turning off the connection for one of the circles over the Riemann surface.  We pick the $\phi_+$ circle.  One also needs to turn off the off-diagonal metric component between the circles.  In the $\mathcal{N}=2$ limit we impose
\begin{equation}
h_{+-} = 0, \qquad *d \partial_+ D_0 =0
\end{equation}  These conditions imply that $D_0$ is of the form
\begin{equation}
D_0 = \widetilde{D}(x,t^-) -\frac{3}{2} s \left(\log s -1\right) + F(t^+).  \label{N2D0}
\end{equation} After plugging into the Monge-Amp\`ere equation \eqref{MAeq}, the only solution with a non-degenerate metric satisfies 

\begin{equation}
a_+ = \frac{2}{3}, \qquad e^{\partial_+ F(t^+)} = \frac{1}{3 a_-^2} \left(c - t^+ \right) \label{sol2}
\end{equation} 

The condition that $a_+ = \frac{2}{3}$ implies $a_- = \frac{1}{3}$ and $\epsilon = \frac{1}{3}$.  This is consistent with the expected value of $\epsilon$ in \eqref{n2p'}  when the line bundle $\mathcal{L}_{p'}$ is trivial.  When we turn off the connection for $\phi_-$ instead, we find $\epsilon= -\frac{1}{3}$ which is consistent with \eqref{n2q'}.

To write the LLM metric, we make the transformation
\begin{equation}
D = \partial_- \widetilde{D}, \qquad y^2 = 4c+2t^-, \qquad y^2 \sin^2(\theta) = 4(c-t^+)
\end{equation} to obtain 
\begin{align}
ds^2_{11} &=H^{-1/3} \left[ds^2_{AdS_5} + \frac{1}{3} H  ds^2_6 \right] \\
ds^2_6 &= (1- y\partial_y D) \left[dy^2 + e^{D} \left(dx_1^2 + dx_2^2\right)\right]  + y^2 d\Omega_2^2 + \frac{4y^2}{- y\partial_y D} \left(d\phi_-+ \frac{1}{2} *d D\right)^2.
\end{align} The coordinates $(\theta, \phi_+)$ combine to a round two-sphere, $d\Omega_2^2$.  The $U_+(1)$ symmetry, dual to $\partial_{\phi_+}$, is enhanced to $SU_+(2)$.    

The warp factor is given as
\begin{equation}
H = \frac{3}{4 y^2} \frac{- y\partial_y D}{1- y\partial_y D}.
\end{equation}  The system is governed by the $SU(\infty)$ Toda equation,
\begin{equation}
\left(\partial_{x_1}^2 + \partial_{x_2}^2 \right) D + \partial_y^2 e^{D} =0.
\end{equation}  This system is LLM as described in GM in \cite{Gaiotto:2009gz}.  

\section{Constant curvature Riemann surface} \label{CCSystems}

In this section, we reduce the system for cases when the surface has constant curvature.  The two-dimensional metric is govern by the Liouville equation and we write it as
\begin{equation}
ds^2\left(\Sigma_g \right) = e^{2A_0(x)} \left(dx_1^2 + dx_2^2 \right), \qquad \left(\partial_{x_1}^2 + \partial_{x_2}^2 \right) A_0 + \kappa e^{2A_0} =0.
\end{equation} The parameter $\kappa$ is the curvature of the surface and takes value in $\{-1,0,1\}$ corresponding to $\mathbb{H}_2$, $T^2$ and $S^2$ respectively.  A genus, $g$, Riemann surface is obtained by replacing $\mathbb{H}_2$ with $\mathbb{H}_2 /\Gamma$ where $\Gamma$ is a Fuchsian subgroup of $PSL(2, \mathbb{R})$, the isometry group of the hyperbolic plane.  In this section, we will restrict to cases when $\kappa^2=1$; the discussion can be straight-forwardly extended to include the torus.  

The conformal factor of the Riemann surface can be written as
\begin{equation}
e^{A_0} = \frac{2}{1+ \kappa \left(x_1^2 + x_2^2 \right)}.  
\end{equation}
It is useful to introduce the one-forms 
\begin{equation}
V = \frac{1}{2(1-g)} * dA_0, \qquad d V = \frac{\kappa}{2(1-g)} e^{2A_0} dx_1 \wedge dx_2.  
\end{equation} It can be written as 
\begin{equation}
V = \frac{\kappa}{1-g} \frac{x_1 dx_2 -x_2 dx_1}{1+ \kappa \left(x_1^2 + x_2^2 \right)}  \label{VCC}
\end{equation} 

The potential, $D_0$, can be written as
\begin{equation}
D_0 =-\kappa A_0(x_1,x_2) \mu_g^2  + F(t^+,t^-), \qquad \mu_g^2  = c_g -(\kappa+z) t^+- (\kappa - z) t^-.
\end{equation}  The equation of motion reduces to
\begin{equation}
\mu_g^2 = s^{5/2} e^{(\partial_+ + \partial_-) F} \left[ \partial_+^2 F \partial_-^2 F - \left(\partial_+ \partial_- F \right)^2 \right]. \label{sepmonamp}
\end{equation}  

The one-forms dual to the $U(1)$s can be written as
\begin{equation}
\eta_+= d\phi_+ - p V, \quad \eta_-= d\phi_+ - q V.  \label{etasCC}
\end{equation} The degrees $(p,q)$ are related to $z$ as 
\begin{equation}
p = \left(\kappa + z\right) \frac{g-1}{\kappa}, \qquad q = \left(\kappa - z\right) \frac{g-1}{\kappa}. \label{pqzCC}
\end{equation}The parameter $z$ is refereed to as twist parameter; $(p-q)$ is the degree of the $U(1)$ bundle over $\Sigma_g$ that is associated to the flavor symmetry $\mathcal{F}$.  Although we were not careful to include the torus in this discussion, we can still continue the final result to include the torus.  We can smoothly take the limit $\kappa\to 0$ in the equation of motion \eqref{sepmonamp} and in the final expression for the one-forms $\eta_\pm$ in \eqref{etasCC}.  

Explicit solutions are obtained by solving the Monge-Amp\`ere equation in \eqref{sepmonamp} for $F(t^+,t^-)$.  The simplest non-trivial ansatz for $F$ yields the $B^3W$ solutions that describe the near horizon region of $M5$-branes wrapped on compact Riemann surfaces \cite{Bah:2011vv,Bah:2012dg}.  We discuss this ansatz and its generalizations next.

\subsection{$B^3W$ Solutions}\label{BBBW system}

One of the simplest class of solutions with constant curvature is obtained from the ansatz
\begin{equation}
F= \frac{3}{4} \mu_s^2 \left(1- \ln \mu_s^2 \right) + \sum_{\alpha=1}^2 \frac{1}{2}\mu^2_\alpha \left( 1-\ln \mu^2_\alpha \right) +  (t_+ + t_-)(\nu+\frac{5}{4} \ln2).
\end{equation} where
\begin{equation}
\mu_s^2 = 2 s, \qquad \mu_1^2 =  c_1 -2 t^+ , \qquad \mu_2^2 = c_2 -2  t^- .
\end{equation} The equation of motion \eqref{sepmonamp} reduces to
\begin{equation}
\mu_g^2 =  e^{2\nu} \left[ 4 \mu_s^2 + 6 \left( a_+^2 \mu_1^2 +  a_-^2 \mu_2^2 \right) \right]. \label{B3Wcons}
\end{equation} 

The equation in \eqref{B3Wcons} leads to algebraic relations which are solved in terms of the twist parameter $z$ and curvature $\kappa$.  We find
\begin{equation}
2 e^{2\nu}= (\kappa^2+3z^2)^{1/2} -\kappa, \qquad \epsilon = \frac{\kappa+ \sqrt{\kappa^2+ 3z^2}}{3z}. \label{evepB3W}
\end{equation} Recall that $a_\pm = \frac{1}{2} (1\pm \epsilon)$.  The parameter $c_g=-2\kappa s_0$ is fixed in terms of $c_{1,2}$, $\kappa$ and $\epsilon$.  The parameters $c_{1,2}$ are not constraint by the equation of motion, we are free to fix them.  

The eleven dimensional metric is 
\begin{align}
ds^2_{11} &= H^{-1/3} \left[ ds^2_{AdS_5} + \frac{1}{3} H \left(\mu_g^2 ds^2\left(\Sigma_g \right) + 3d\mu_s^2 + 2 \sum_{\alpha=1}^2 \left(d\mu_\alpha^2 +\mu_\alpha^2 \eta_\alpha^2 \right) \right) \right] \label{B3W}
\end{align} with the constraint
\begin{equation}
 \mu_s^2 + a_+  \mu_1^2 +  a_-  \mu_2^2 = a_+  c_1 + a_-  c_2.  \label{ellipsoid}  
\end{equation} The warp factor is
\begin{equation}
H = \frac{ e^{2\nu}}{\mu_g^2} =  \frac{ 1 }{ 4 \mu_s^2 + 6a_+^2  \mu_1^2 + 6 a_-^2  \mu_2^2 }.  
\end{equation} The warp factor is positive and finite when $a_+ c_1 +a_- c_2>0$.  This is the only condition that can be imposed on the $c$'s, we can fix the combination $a_+ c_1 +a_- c_2=1$ without lost of generality.  The metric is smooth in the regions where $\mu_\alpha=0$ when the periods of $\phi_\pm$ are $\Delta \phi_\pm =2\pi$.  Quantization of the first chern number of the $U(1)$ bundles over the Riemann Surface implies that $p$ and $q$ are integers.    

The constraint in \eqref{ellipsoid} describes a two dimensional surface in $\mathbb{R}^3$.  When $\kappa=-1$, $\epsilon$ is bounded as $-1 \leq \sqrt{3} \epsilon \leq 1$ for all $z$; this implies that $a_\pm >0$.  This bound guarantees that the surface described in \eqref{ellipsoid} is always compact.  When $\kappa =1$, $|\epsilon|\leq 1$ if and only if $|z| \geq 1$, in which case the surface described by \eqref{ellipsoid} is compact.  When $\kappa =1$ and $|z|<1$, the four manifold normal to the Riemann surface is non-compact.  It is interesting to wonder whether it can be compactified by moding with a discreet subgroup.  This will require a more detailed understand of the four manifold and will be investigated in future works.

\subsection{Generalization of $B^3W$ solutions}

We can straight-forwardly generalize the $B^3W$ ansatz to
\begin{align}
F &= \frac{3}{4} \mu_s^2 \left(1- \ln \mu_s^2 \right) +  (t_+ + t_-)(\nu+\frac{5}{4} \ln2) \nonumber \\
&+ \frac{n_u}{4} \mu_u^2 \left(1-\ln \mu_u^2\right) + \frac{1}{4}\sum_{\alpha=1}^k n_\alpha \mu^2_\alpha \left( 1-\ln \mu^2_\alpha \right) \label{ansatzCC}
\end{align} where 
\begin{equation}
\mu^2_\alpha = c_\alpha - (1 +r_\alpha) t^+ - (1 -r_\alpha)  t^-, \qquad \mu_u^2= c_u+2(t_+-t_-).    
\end{equation} Generically, there are $k$ independent $\mu_\alpha$'s, each of which defined by two parameters $c_\alpha$ and $r_\alpha$.  The parameter, $n_\alpha$, is the multiplicity of $\mu_\alpha$.  The $B^3W$ ansatz is recovered when $k=2$, $n_\alpha =2$, $n_u=0$,  and $r_1=-r_2=1$.

The metric for the ansatz in \eqref{ansatzCC} is 
\begin{align}
ds^2_{11} &= H^{-1/3} \left[ ds^2_{AdS_5} + \frac{1}{3} H \left( e^{2\nu} \det(g) \mu_s^{2} \prod_{\alpha=1}^k \mu_\alpha^{n_\alpha} \; ds^2\left(\Sigma_g\right)+ds^2_{4} \right) \right] \label{metric11CC} \\
ds^2_4 &= 3d\mu_s^2 +n_u \left(d\mu_u^2 + \frac{4}{\mu_u^2 \det(h)} \eta_{u \perp}^2 \right) + \sum_{\alpha=1}^k n_\alpha \left(d\mu_\alpha^2 + \frac{4}{\mu_\alpha^2 \det(h)} \eta_{\alpha \perp}^2 \right) \label{metric4CC}
\end{align} where the circles are given as
\begin{equation}
\eta_{u\perp} = \eta_+ + \eta_-, \qquad \eta_{\alpha \perp} = \frac{1 +r_\alpha}{2} \eta_- - \frac{1 -r_\alpha}{2} \eta_+.  
\end{equation}

The metric in the \eqref{metric4CC} is on $\mathbb{R} \times\left(\mathbb{R} \times S^1\right)^{k+1}$.  The first $\mathbb{R}$ factor is coordinatized by $\mu_s$.  Each $\mu_\alpha$ can be identified with the radius of a plane $(\mathbb{R} \times S^1)$ whose circle is given as $\eta_\alpha =   \frac{1 +r_\alpha}{2} \eta_+ + \frac{1 -r_\alpha}{2} \eta_-$.  The four dimensional space in the eleven dimensional metric \eqref{metric11CC} is obtained by restricting the metric \eqref{metric4CC} to the surface given by the equations
\begin{align}
(r_\alpha -\epsilon) \mu_u^2 + \mu_s^2 + \mu_\alpha^2 &= R_{\alpha} 
\end{align} where the $R$'s are related to the $c$'s.  Using these constraints, one can derive a similar relation between any three $\mu$'s.  This is particularly important since we can turn off a given $\mu_\alpha$ by fixing $n_\alpha=0$.  In such case, one writes a set a equations that do not involve $\mu_\alpha$.  A similar relation can be written for the coordinates on the circles:
\begin{equation}
(\epsilon-r_\alpha) \eta_{u \perp} + 2 \eta_{s\perp} + 2 \eta_{\alpha \perp} =0, \qquad \eta_{s\perp} = a_- \eta_+ - a_+ \eta_-.  
\end{equation}

 The determinants are given as
\begin{align}
\qquad \det(h) &= \frac{1}{2} \sum_{\alpha,\beta} \frac{ n_\alpha n_\beta (r_\alpha-r_\beta)^2}{4\mu_\alpha^2 \mu_\beta^2}+  \sum_\alpha \frac{n_\alpha n_u }{\mu_\alpha^2 \mu_u^2} , \label{dethCC}\\
 \det(g) &= \det(h) + \frac{3}{\mu_s^2} \left[\frac{n_u}{\mu_u^2} + \frac{1}{4}\sum_\alpha \frac{n_\alpha (\epsilon-r_\alpha)^2}{\mu_\alpha^2}  \right]. \label{detgCC}
\end{align}  From these expressions one can compute the warp factor
\begin{equation}
H = \frac{1}{4 \mu_s^2} \frac{\det(h)}{\det(g)}.  \label{warpHCC}
\end{equation}

We can associate a killing vector $\ell_\alpha$ to every $\mu_\alpha$ defined as
\begin{align}
\ell_u &= \frac{1}{2}  \frac{\partial}{\partial \phi_+} - \frac{1}{2} \frac{\partial}{\partial \phi_-} \\
\ell_s &= a_+ \frac{\partial}{\partial \phi_+} + a_- \frac{\partial}{\partial \phi_-}  \\
\ell_\alpha &= \frac{1+r_\alpha}{2} \frac{\partial}{\partial \phi_+} + \frac{1-r_\alpha}{2} \frac{\partial}{\partial \phi_-}.
\end{align} The killing vector $\ell_u$ and $\ell_s$ are respectively dual to the flavor symmetry, $\mathcal{F}$, and the superconformal R-symmetry, $R_{\mathcal{N}=1}$, in field theory.

The norms of the killing vectors are such that as
\begin{align}
\mu_u \; &\to \; 0 , \qquad ||\ell_u||^2  \; \rightarrow \; \frac{\mu_u^2}{n_u} \\
\mu_\beta \; &\to \; 0 , \qquad ||\ell_\beta||^2  \; \rightarrow \; \frac{4\mu_\beta^2}{n_\beta}.  
\end{align} This implies that the circle dual to the killing vector, $\ell_\alpha$, shrinks in the region $\mu_\alpha=0$.  Since each such circle is generically non-trivially fibered on the Riemann surface, flux quantization of the connection leads to quantization conditions on the $r_\alpha$ parameters.  From the norm of $\ell_s$, one can see that its orbits never shrink unless $r_\alpha = \epsilon$ for some $\alpha$.  

The solution to $\mu_\alpha=0$ can be represented as a line on the $(t^+,t^-)$ plane with slope, $m_\alpha = -\frac{1-r_\alpha}{1+r_\alpha}$.  Since a circle shrinks along every such line, the regions where the $\mu$'s vanish can be identified with the boundaries of the space.  When two different $\mu_\alpha =0$ lines intersect, the full $(\phi_+,\phi_-)$ torus shrinks; this leads to an orbifold singularity at the intersection.  This hints at the presence of certain sources in these regions.  The field theories dual to these solutions will have global symmetries associated to these orbifold singularities.  It interesting to understand these sources in these solutions; this will require a better understand and control of the four form flux.  We leave these questions to future studies of the solutions.  

In order to avoid curvature singularities, the warp factor $H$ must stay finite and positive everywhere.  The determinants in \eqref{dethCC} and \eqref{detgCC} have the same pole structure in the regions $\mu_\alpha=0$ and $\mu_u=0$, this guarantees the finiteness of $H$ in \eqref{warpHCC}.  The determinant $\det(g)$ also have a pole at $\mu_s^2 =0$ while $\det(h)$ does not, this is cancelled in the expression of $H$ by the overall $\frac{1}{\mu_\alpha^2}$ in \eqref{warpHCC}.

The Riemann surface shrinks in the region $\mu_\alpha \to 0$ when $n_s \geq 2$; and when $n_s \leq 2$, its radius blows up.  It is natural to fix $n_\alpha =2$ in order to have a finite Riemann surface.  
    
Supersymmetric solutions are obtained by solving \eqref{sepmonamp} which reduces to 
\begin{equation}
\mu_g^2 = e^{2\nu}\det(g) \mu_s^2 \prod_{\alpha =1}^k \mu_\alpha^{n_\alpha} . \label{mugCC} 
\end{equation}
This equation reduces to algebraic relations between parameters when we expand in powers of $(t^+,t^-)$.  When $k=0$, $\det(h)$ in \eqref{dethCC} vanishes.  The first solutions to the equations are found when $k=2$.

\paragraph{$k=2$ solutions}
When $k=2$ and $n_1=n_2 =2$, the system can be solved explicitly.  The solution can written in terms of $z_0$ and $\epsilon_0$ which are related to the $z$ and $\epsilon$ as
\begin{equation}
z =\frac{r_1 +r_2}{2}\kappa + \frac{r_1-r_2}{2} z_0 \qquad \epsilon= \frac{r_1 +r_2}{2}+ \frac{r_1-r_2}{2} \epsilon_0.
\end{equation} We find
\begin{equation}
 2e^{2\nu} = \frac{4}{(r_1 -r_2)^2} \left(\sqrt{\kappa^2 + 3z_0^2 }-\kappa \right), \qquad \epsilon_0 = \frac{\kappa + \sqrt{\kappa^2 + 3 z_0^2}}{3z_0^2}.  
\end{equation}  When $r_1=-r_2 =1$, the solution reduces to $B^3W$ systems.  When the $r$'s are arbitrary, the solutions are transformation of $B^3W$ under the $S$-symmetry described in equation \eqref{symmetry}.  The matrix $A$ is given as
\begin{equation}
\left(\begin{array}{c} \eta_+ \\ \eta_- \end{array} \right) = \frac{1}{2} \left(\begin{array}{cc} 1+r_1 & 1+r_2 \\ 1-r_1 & 1-r_2 \end{array} \right) \left(\begin{array}{c} \eta_+^0 \\ \eta_-^0 \end{array} \right)
\end{equation} where $\eta_\pm^0$ are the one-forms that appear in $B^3W$.

\paragraph*{General $k$}

For $k>2$, the supersymmetry equation in \eqref{mugCC} cannot be solved everywhere.  We can solve the equation only regions where two $\mu_\alpha=0$ lines intersect.  This suggest that the ansatz in \eqref{ansatzCC} is not general enough to fully capture the system.  The main problem is that the boundary of the space on $(t^+,t^-)$ is piece-wise defined and the physics of the intersections when $k\geq3$ is poorly understand.  This may indicate presents of sources or may correspond to orbifolds of the $(2,0)$ SCFT wrapped on a Riemann surface.  In order to fully capture these systems, one may need to perform a B\"acklund transformation of the Monge-Amp\`ere to new coordinates.  In the transformed system, the coordinates $t^\pm$ are multivalued functions.  It is yet unclear to the author on how to proceed.  

Although we are unable to solve the supersymmetry constraint, it may be that the metric in \eqref{metric11CC} solves the M-theory Einstein equations.  This would lead to a new class of non-supersymmetry $AdS_5$ solutions in M-theory.  A better control of the four-form flux is required to  check this.

\section{Punctures} \label{punc}

The main goal of this paper is understand the $AdS_5$ solutions obtained by wrapping $M5$-branes on Riemann surfaces with punctures.  In this section  we study punctures. They appear in gravity as localized sources on Riemann surfaces.  Throughout this section, we use the terms sources and punctures synonymously. 

Before we study the general system of $M5$-branes discussed above, we first review sources on two-dimensional Riemann surfaces.   These are governed by the Liouville system:
\begin{equation}
ds^2 = e^{2A_0} \left(dx_1^2 + dx_2^2\right), \qquad \mbox{with} \qquad \left(\partial_{x_1}^2 + \partial_{x_2}^2\right)  A_0 = - \kappa e^{2A_0}.
\end{equation} The compact solutions of this system are discussed in section \ref{CCSystems}.  In presences of punctures, the Liouville equation is modified with delta functions sources.  In the region near a puncture, there is a shrinking circle  and the Liouville equation can be written as
\begin{equation}
\frac{1}{r} \partial_r\left( r \partial_r A_0\right) = - \kappa e^{2A_0} + 2\pi (k-1) \delta^2(r) \label{2Dpunc}
\end{equation}  where $(x_1= r\cos(\varphi), x_2 =r \sin(\varphi))$.  There are $(k-1)$ units of sources at $r=0$\footnote{Our normalization for the delta function is 
\begin{equation} \int \delta^2(r-r_0) rdr d\varphi =1, \qquad \mbox{thus} \qquad \delta^2(r-r_0) = \frac{1}{2\pi r} \delta(r-r_0). \end{equation}}.  The solution to $A_0$ has two parts, a logarithmic term due to the puncture and a background term due to the curvature $\kappa$:
\begin{equation}
A_0 = (k-1) \ln r + \widetilde{A}_0(r), \qquad \mbox{with} \qquad e^{2\widetilde{A}_0} = \frac{4k^2}{\left(1+ \kappa r^{2k}\right)^2}.
\end{equation} The metric near the puncture region can be written as
\begin{equation}
ds^2 = \frac{4}{\left(1+\kappa \rho^2 \right)^2} \left( d\rho^2 + k^2 \rho^2 d\varphi^2 \right) \label{2Dorbi}
\end{equation}  where $\rho = r^k$.  the circle $\varphi$ shrinks at $\rho=0$ with a period $\frac{2\pi}{k}$.  In regions away from the puncture, the period of $\varphi$ is $2\pi$, this is necessary for the identification $(x_1= r\cos(\varphi), x_2 =r \sin(\varphi))$.  The presences of the sources leads to a conical deficit and therefore a  $\mathbb{Z}_k$ orbifold fixed point at $\rho=0$.  This is precisely the manifestation of a puncture of the two-dimensional surface.  

The spectrum of punctures in class $S$ is given by the different ways we can embed the orbifold fixed points from \eqref{2Dpunc} and \eqref{2Dorbi}, in the system described in \ref{Gravitydual}.  These sources are localized on the $(x_1,x_2)$ plane in the metric \eqref{metricgen}.  In the class $\mathcal{S}$ systems, the Liouville equation is generalized to the modified Monge-Amp\`ere equation for $D_0$, \eqref{MAeq}.  Therefore the sources for $A_0$ in \eqref{2Dpunc} generalize to sources of $D_0$.  The most general expression we can write to include these sources is 
\begin{equation}
\left(\partial_{x_1}^2 + \partial_{x_2}^2\right) D_0 = e^{2A} + \sum_\alpha P_\alpha(t^+,t^-) \delta(x_1 -x_1^\alpha) \delta(x_2 -x_2^\alpha)
\end{equation} where
\begin{equation}
e^{2A} =s^{5/2} e^{(\partial_+ + \partial_-)D_0} \left[\partial_+^2 D_0 \partial_-^2 D_0 - \left(\partial_+ \partial_- D_0\right)^2 \right].
\end{equation}
The index $\alpha$ labels the sources located at $(x_1^\alpha, x_2^\alpha)$.  A puncture can have a non-trivial profile in the directions normal to the Riemann surface; this is determined by the functions $P_\alpha$.  To classify the punctures, we need to find the set of functions $P_\alpha$ for which there exist solutions of $D_0$ that lead to a regular metric in the neighborhood of the source.  It is important to note that the functions $P_\alpha$ are not required to be smooth on the $(t^+,t^-)$ plane.  

In the neigborhood of a puncture, the metric on the $(x_1,x_2)$ plane can be written as
\begin{equation}
e^{2A(r,t^+,t^-)} \left(dr^2+ r^2 \varphi^2 \right)
\end{equation} where $(x_1= r\cos(\varphi), x_2 =r \sin(\varphi))$.  The circle $\varphi$, which appears in the neigborhood of the puncture, must have period $2\pi$.   
The supersymmetry equation, for the puncture with profile $P$, is
\begin{equation}
\frac{1}{r} \partial_r\left( r \partial_r D_0 \right)= e^{2A} + 2\pi P(t^+,t^-) \delta^2(r).  \label{punctureseq}
\end{equation}  When $P$ is a smooth function, the potential $D_0$ can be written as
\begin{equation}
D_0 = P(t^+,t^-) \ln r + \widetilde{D}(r,t^+,t^-). \label{D0fexp}
\end{equation} After plugging $D_0$ in equation \eqref{punctureseq} and expand in powers of $\ln(r)$, we obtain 
\begin{align}
0 &= \partial_+^2 P \partial_-^2 P - \left(\partial_+ \partial_- P \right)^2 \label{punc1}\\
0 &= \partial_+^2 P \partial_-^2 \widetilde{D} + \partial_-^2 P \partial_+^2 \widetilde{D} - 2 \partial_+ \partial_- P \partial_+ \partial_- \widetilde{D} \label{punc2}\\
\frac{1}{r} \partial_r\left( r \partial_r \widetilde{D} \right) &= s^{5/2} r^{(\partial_+ + \partial_-)P} e^{(\partial_+ + \partial_-)\widetilde{D}} \left[\partial_+^2 \widetilde{D} \partial_-^2 \widetilde{D}  - \left(\partial_+ \partial_- \widetilde{D} \right)^2 \right].  \label{punc3}
\end{align} We refer to these equations as the puncture equations.  The first one determines the choices for $P$, the last two are compatibility conditions with the background surface.  The equation in \eqref{punc3} makes sense as a single equation when $(\partial_+ + \partial_-)P$ is differentiable.  Otherwise the equation would have to be split into different parts and sources would be needed for $\widetilde{D}$ at the junctions.  This implies that the expansion in \eqref{D0fexp} is not valid when $(\partial_+ + \partial_-)P$ is not differentiable.  

We will not aim to find the most general solution or form of $P$ in this paper.  The Profile $P$ can be fixed to a constant.  In this case, the first two puncture equations are trivially solved and the last reduces to the original system. 
The simplest non-trivial examples are cases when $P$ is linear in the $t$'s.  We focus on such examples for the rest of the paper.    

\subsection*{Linear Profile}

Consider a puncture profile
\begin{equation}
P= n_+ t^+ + n_-t^-.  
\end{equation} The first and second puncture equations, \eqref{punc1} and \eqref{punc2}, are trivially solved.  The last puncture equation, \eqref{punc3}, yields an equation for $\widetilde{D}$ that can be written as
\begin{equation}
\frac{k^2}{\rho} \partial_\rho\left( \rho \partial_\rho \widetilde{D} \right) = s^{5/2} e^{(\partial_+ + \partial_-)\widetilde{D}} \left[\partial_+^2 \widetilde{D} \partial_-^2 \widetilde{D}  - \left(\partial_+ \partial_- \widetilde{D} \right)^2 \right]
\end{equation} where $2(k-1)=n_++n_-$ and $\rho= r^{k}$.  In the special case when $n_+ =-n_-$, the puncture is completely smoothed out since $k=1$.  In this description, $k<<N$ where $N$ is the number of $M5$-branes wrapping the Riemann surface.

The equation for $\widetilde{D}$, as a function of $\rho$, is the same as the original system but without the localized sources.  To describe this region we consider the $B^3W$ ansatz for $\widetilde{D}$:
\begin{align}
\widetilde{D} &= -\kappa \widetilde{A}_0(\rho) \mu_g^2 \nonumber \\
&+ \frac{3}{4} \mu_s^2 \left(1-\ln \mu_s^2\right) + \frac{1}{2} \sum_{\alpha=1}^2 \mu_\alpha^2 \left(1-\ln \mu_\alpha^2 \right) + (t^+ +t^-) \left(\nu + \frac{5}{4} \ln 2 \right).
\end{align} The $\mu$s are defined in section \ref{BBBW system}.  The solution for $\epsilon$ and $e^{2\nu}$ are given in equation \eqref{evepB3W}.  The eleven-dimensional metric is
\begin{align}
ds^2_{11} &= H^{-1/3} \left[ ds^2_{AdS_5} + \frac{1}{3} H \left(\mu_g^2 \frac{4}{(1+\kappa \rho^2)^2} \left(d\rho^2 + k^2 \rho^2 d\varphi^2 \right) + ds^2_4 \right) \right]  \\
ds^2_4 &= 3d\mu_s^2 + 2 \left(d\mu_1^2 +\mu_1^2 \left(d\phi_+ -\frac{n_+}{2} d\varphi-p \widetilde{V} \right)^2\right)  \nonumber\\
&+ 2 \left(d\mu_2^2 +\mu_2^2 \left(d\phi_- -\frac{n_-}{2} d\varphi-q \widetilde{V} \right)^2 \right)
\end{align} where
\begin{equation}
\widetilde{V} = \frac{1}{2(1-g)} *d\widetilde{A}_0 = \frac{\kappa}{1-g} \frac{k \rho^2}{1+\kappa \rho^2} d\varphi.  
\end{equation} In the region near $\rho=0$, the internal metric can be written as
\begin{align}
ds^2_6&= 4\mu_g^2 \left(d\rho^2 + \frac{4 k^2}{(b_-n_-+b_+ n_+)^2} \rho^2 d\phi_b^2\right)+ 3 d\mu_s^2 + 2 d\mu_1^2 + 2 d\mu_2^2 \\
&+ \frac{\mu_1^2 n_+^2 + \mu_2^2 n_-^2}{2} \left(d\varphi - 2 \frac{n_+ \mu_1^2 d\phi_+ + n_- \mu_2^2 d\phi_-}{\mu_1^2 n_+^2 + \mu_2^2 n_-^2} \right)^2 + \frac{2 \mu_1^2 \mu_2^2 }{\mu_1^2 n_+^2 + \mu_2^2 n_-^2} d\phi_u^2 
\end{align} where
\begin{equation}
\phi_b = b_+ \phi_+ + b_- \phi_-, \qquad \phi_u = n_- \phi_+ -n_+ \phi_-.  
\end{equation} The only restriction for the possible choices on the $b$'s is that $b_- n_- +b_+ n_+ \neq 0$; this is a necessary condition for $\phi_b$ and $\phi_u$ to be independent.  The geometry is smooth when the period of $\phi_b$ is $2\pi$ and $b_- n_- +b_+ n_+ = \pm 2k$.  

The main observation is that orbifold fixed points with linear profiles $P$ are smoothed out by the bundle structure in the full eleven-dimensional geometry.  The physical parameters $\epsilon$ and $e^{2\nu}$ do not have any dependence on the puncture data.  However the volume of the internal manifold will have a correction due to $k$.  This may lead to correction to the central charge of the dual theory which would be suppressed by $N$, the number of $M5$-branes wrapping the surface.  At finite $N$, $\epsilon$ and $e^{2\nu}$ are expected to pick up corrections.  In order to check this, we need to work out the flux quantization.  We address these issues in future explorations when the four form flux is computed.    

\section{Punctures and $D6$-branes} \label{puncD6}

In this section, we study punctures with piece-linear profiles.  The simplest example is
\begin{equation}
P= \left( c_1 + 2n_1 t \right) \theta(t_0-t) +  \left(c_2 + 2n_2 t \right) \theta(t-t_0) \label{piecewise}
\end{equation} where $\theta(t_0-t)$ is the step function. 

 The coordinate $t$ is given as
\begin{equation}
t= n_+ t^+ + n_- t^-.
\end{equation}  The $n$'s can be restricted  $n_+ + n_- =1$ or $n_+ =-n_- =1$ without lost of generality.  In this example the profile $P$ is a continuous linear function of $t$ with slope $2n_1$ when $t<t_0$ and $2n_2$ when $t>t_0$.  In regions away from $t=0$, the puncture is smoothed out since the profile linear.  Our interest here is to understand what is happening at the intersection, $t=t_0$ and $r=0$.  

It is instructive to see how the expansion in \eqref{D0fexp} fails when we plug the piece-wise $P$, in \eqref{piecewise}, into equations \eqref{punc1}-\eqref{punc3}.  The profile solves the first puncture equation \eqref{punc1}.  However, in the last puncture equation, \eqref{punc3}, there is a discontinuity in the right hand side due to the $r^{(\partial_+ + \partial_-)P}$.  This implies a source for $\widetilde{D}$ at $t=t_0$.  The second puncture equation \eqref{punc2} reduces to
\begin{equation}
(n_2-n_1) \delta(t-t_0) \left(n_+ \partial_- - n_-\partial_+\right)^2 \widetilde{D} = 0.
\end{equation} These conditions suggest that near $t=t_0$ the potential $D_0$ should be expanded as
\begin{equation}
D_0 = \widetilde{D}(t,r) + F(t^+,t^-)+ \dots \label{D0step}
\end{equation} The ellipsis correspond to corrections to $D_0$ in regions away from the puncture.  The potential $\widetilde{D}$ will capture the source and the function $F$ will capture the behavior of the metric in the neighborhood of the puncture.  We reiterate the point that the solutions obtain here yields the boundary conditions of $D_0$ associated to the step function source $P$ in \eqref{piecewise}.  

In appendix \ref{GenLLM}, the general Monge-Amp\`ere system is reduced with the exact ansatz \eqref{D0texact}.  The results in appendix \ref{GenLLM} can be used to describe the $t=t_0$ region of the piece-wise profile $P$ \eqref{piecewise}. 

The general equation for $\widetilde{D}$ in appendix \ref{GenLLM} is \eqref{GenLLMeq}.  Near the source we write it as
\begin{equation}
\frac{1}{r} \partial_r \left(r\partial_r \widetilde{D} \right) = \left(h_0(t) - h_1(t) \partial_t^2 \widetilde{D} \right) e^{(n_+ + n_-) \partial_t \widetilde{D}} +2\pi P(t) \delta^2(r).  \label{stepeq}
\end{equation} There are different solutions for $F$ and the $h$'s for when $n_+=-n_- =1$, $n_\pm = a_\pm$ and for when $n_+ + n_-=1$.  In each of these cases, the equation for $\widetilde{D}$ reduces differently and therefore leads to different types of sources at $t=t_0$.  We study them independently.

\subsection{$\mathcal{N}=2$ Punctures and $D6$ branes}\label{N2D6}

The metric when $n_++n_-=1$ is described in section \ref{N2LLM} and given in equation \eqref{N2LLMmet}.  In this section, we discuss the punctures when $c_1=0$ where the metric reduces to LLM and preserves $\mathcal{N}=2$ supersymmetry.  We return to cases with non-vanishing $c_1$ in section \eqref{N1M9}. 

The punctures in LLM were studied and described in \cite{Gaiotto:2009gz}.  We will review their description and discuss their $D6$ brane interpretation.  

When $c_1=0$, the metric  in \eqref{N2LLMmet} is non-degenerate only when $c$ is non-vanishing, we can fix it as $c=-1$ without lost of generality.  The LLM metric is recovered by making the coordinate transformation $y^2 = 1+2t$ and the field redefinition $e^D = \frac{h_1}{1-2ct}e^{\partial_t \widetilde{D}_0}$.  Equation \eqref{stepeq} becomes
\begin{equation}
\frac{1}{r} \partial_r \left(r \partial_r D\right) + \partial_y^2 e^D = 2\pi \left(2n_1 \theta(y_0-y) + 2n_2 \theta(y-y_0) \right) \delta^2(r)
\end{equation} where $y_0^2 = 1+ 2t_0$. 

The potential $D$ satisfies an axially symmetric $SU(\infty)$ Toda equation; its sources can be studied by performing the b\"acklund transformation \cite{Ward:1990qt,Lin:2004nb}
\begin{equation}
y= \rho \partial_\rho V, \qquad \ln(r) = \partial_z V, \qquad \rho^2 = r^2 e^D. \label{backToda}
\end{equation} Under this transformation, the $SU(\infty)$ Toda equation is satisfied when $V$ solves 
\begin{equation}
\frac{1}{\rho} \partial_\rho \left(\rho \partial_\rho V\right) + \partial_z^2 V =0.
\end{equation}  
Next we need to understand how the profile $P$ appears as a source for $V$.  The region $r=0$ maps to the region $\rho =0$.  In regions away from the source at $y=y_0$, the potential can be approximated as $e^D=r^{2n_1}$ for $y<y_0$ and $e^D=r^{2n_2}$ for $y>y_0$.  Under the B\"acklund transformation, $V$ is of the form $V= y_1(z) \ln(\rho)$ for $y<y_0$  and $V= y_2(z) \ln(\rho)$ for $y>y_0$ with 
\begin{equation}
y_1(z) = b_1 + \frac{1}{1+n_1}z, \qquad y_2(z) = b_2 + \frac{1}{1+n_2}z
\end{equation} where the $b$'s are constants.  The point $y_0$ maps to $z_0$ where $y_1(z_0)=y_2(z_0)=y_0$.  The profile $P$ maps to 
\begin{equation}
\frac{1}{\rho} \partial_\rho \left(\rho \partial_\rho V\right) + \partial_z^2 V =  2\pi \left(y_1(z) \theta(z_0-z) + y_2(z) \theta(z-z_0) \right) \delta^2(\rho).  \label{Vsource}
\end{equation} 

The metric after the transformation is 
\begin{align}
ds^2_{11} &= H^{-1/3} \left[ds^2_{AdS_5} + \frac{1}{3} H ds^2_6 \right] \\
ds^2_6 &= \Delta \left(d\rho^2 + dz^2 + \frac{2\dot{V}\rho^2}{2\dot{V} -\ddot{V}}d\hat{\chi}^2 \right) + \frac{2\dot{V}-\ddot{V}}{V''} \left(d\varphi - \frac{2\dot{V}'\dot{V}}{2\dot{V}-\ddot{V}} d\hat{\chi} \right)^2 + \dot{V}^2 d\Omega^2 
\end{align} where $\dot{V}= \rho \partial_\rho V$ and $V' = \partial_z V$.  The $\chi$ circle is shifted as $\hat{\chi} = \chi + \varphi$.  The warp factors are
\begin{equation}
H = \frac{3V''}{2 \dot{V}\Delta}, \qquad \Delta = \dot{V}'^2+ V''(2\dot{V} -\ddot{V}).  
\end{equation}  From the equation of $V$, we observe that $V''$ has a point source at $(\rho=0,z=z_0)$, it is given as
\begin{equation}
V'' = \frac{k}{2}  \frac{1}{\sqrt{\rho^2 + (z-z_0)^2}}, \qquad k =\left(\frac{1}{1+n_1} - \frac{1}{1+n_2}\right).
\end{equation}
 In the region near the source, the metric is 
\begin{equation}
ds^2_{11} =H^{-1/3}\left[ds^2_{AdS_5} + \frac{V''}{2y_0} (dz^2 + d\rho^2 + \rho^2 d\hat{\chi}^2 ) + \frac{1}{2y_0V''} \left(d\varphi- \dot{V}' d\hat{\chi}\right)^2 + \frac{1}{4}d\Omega^2 \right].  \label{nearmet}
\end{equation}  The overall warp factor goes to a constant, $H=\frac{3}{4y_0^2}$.  

The metric is positive definite when $k>0$ $(n_2>n_1)$.  We make the coordinate transformation $\left(\rho =\frac{1}{2} R^2 \sin(\theta), z =\frac{1}{2} R^2 \cos(\theta)\right)$ and write the $(\rho,z,\hat{\chi},\varphi)$ directions as
\begin{equation}
ds^2_{\mathbb{R}/\mathbb{Z}_k} = dR^2 + R^2 \left[\frac{1}{4} \left(d\theta^2 + \sin^2(\theta) d\hat{\chi}^2 \right) + \frac{1}{k^2} \left(d\varphi - k \sin^2(\theta/2) d\hat{\chi}\right)^2\right].
\end{equation} The parameter $k$ is the degree of a $U(1)$ bundle over $S^2$.  It is quantized and takes value in $k \in \mathbb{Z}_+$.  When $k=1$, the space is $\mathbb{R}^4$ and $\mathbb{R}^4/\mathbb{Z}_k$ in general.
The internal six-dimensional manifold in \eqref{nearmet} reduces to a simple product of a round $S^2$ with $\mathbb{R}^4/\mathbb{Z}_k$.  This orbifold fixed point leads to an $SU(k)$ gauge symmetry in $AdS_5$ and therefore a $SU(k)$ global symmetry in the field theory.  

In the region near the source, the $\varphi$ circle is shrinking and we can interpret it as a shrinking M-theory circle.  The metric then describes $k$ D6-branes wrapped on $AdS_5 \times S^2$.  The $SU(k)$ gauge symmetry in $AdS$ is due to the gauge field on the world volume of the $D6$-branes.  The R-symmetry in this region is $SU(2) \times U(1) $; the $SU(2)$ corresponds to the isometries of the two-sphere wrapped by the $D6$-branes while the $U(1)$ corresponds to the generator dual to the $\chi$ circle.  This symmetry that appears near the source is an enhancement of the global $U_+(1)\times U_-(1)$ symmetry that underlie the general system.  The generator of the $U(1)$ is $n_+ \frac{\partial}{\partial \phi_+}+ n_- \frac{\partial}{\partial \phi_-}$.  The orthogonal combination enhances to the $SU(2)$ symmetry.  

Since the potential $V$ satisfies the Laplace equation everywhere, one can consider a collection of these $D6$-branes at along the line $t$.  This yield the general $\mathcal{N}=2$ puncture with the continuous piece-wise linear profile
\begin{equation}
P(t) = \sum_i (c_i+2n_i t) \theta(t_i -t)\theta(t_{i+1} -t). \label{Ptableaux}
\end{equation} The function $P$ has slope $n_i$ in the interval $(t_i,t_{i+1})$ and the system have $k_i$ $D_6$-branes at $t=t_i$ where
\begin{equation}
k_i = \frac{1}{1+n_{i}} - \frac{1}{1+n_{i+1}}.
\end{equation} These are the $\mathcal{N}=2$ punctures described by Gaiotto Maldacena in \cite{Gaiotto:2009gz}.

 From $\mathcal{N}=1$ point of view, the punctures with profile $P(t)$ in \eqref{Ptableaux} have an additional label, $(n_+,n_-)$.  These parameters determine the R-symmetry of the $\mathcal{N}=2$ system near the source.  At the location of the $D6$-branes, the orbits of the killing vector $n_+ \frac{\partial}{\partial \phi_+}+ n_- \frac{\partial}{\partial \phi_-}$ shrink.  Globally the periods of $\phi_\pm$ can be fixed to $2\pi$.  This implies that the parameters $n_\pm$ must be rational.  We can then write them as 
\begin{equation}
n_+ = \frac{p}{p+q}, \qquad n_- = \frac{q}{p+q}
\end{equation}  where $p$ and $q$ are co-prime integers.  The $D6$-branes can be labelled with $(p,q)$.  Two punctures with different choices of $(p,q)$ will each preserve a $U(1)\times SU(2)$ R-symmetry that corresponds to different enhancements of the $U_+(1) \times U_-(1)$ symmetries which is preserved everywhere.  
 
 An $\mathcal{N}=1$ puncture can be obtained from the $\mathcal{N}=2$ punctures by superpositing different $(p,q)$ profiles at the same point on the Riemann surface.  The description of these sources is beyond the ansatz in \ref{stepeq}.  The $D6$-branes with different $(p,q)$ will trace non-parallel line on the $(t^+,t^-)$ plane and therefore will intersect.  The ansatz in \ref{stepeq} cannot capture the physics at the intersection, one needs to solve the equation in \eqref{MAeq}.  In this region also, the $(\phi_+, \phi_-)$ torus shrink.  A generalization of the B\"acklund transformation in \eqref{backToda} for the Monge-Amp\`ere system \eqref{MAeq} is required.  We continue to explore how to do this.

\subsection{$\mathcal{N}=1$ Punctures and $D6$-branes}\label{N1D6}

In this section we discuss the piece-wise punctures on the metric in \eqref{n=aMet}; these are cases when $n_\pm =a_\pm$ in \eqref{piecewise}.  Equation \eqref{stepeq} reduces to equation \eqref{n=aLLM} with the $P(t)$ source: 
\begin{equation}
\frac{1}{r} \partial_r\left(r \partial_r D \right) + \partial_y^2 e^D - \frac{1}{y} \partial_y \left[ \frac{2}{1-cy^2} e^D\right] = 2\pi \left(2n_1 \theta(y_0-y) + 2n_2 \theta(y-y_0) \right) \delta^2(r).  \label{n=atoda}
\end{equation} In the region near the source, the term $\frac{1}{y} \partial_y \left[ \frac{2}{1-cy^2} e^D\right] $ is subleading when compared to the first two.  In this approximation the $D$ potential satisfied the $SU(\infty)$ Toda equation in the neighborhood of the puncture.  We can then use the B\"acklund transformation in \eqref{backToda} and \eqref{Vsource}.  The metric near the source can be written as
\begin{align}
ds^2_{11} &=H^{-1/3}\left[ds^2_{AdS_5} + \frac{V''}{6y_0} (dz^2 + d\rho^2 + \rho^2 \eta_0^2 ) + \frac{1}{6y_0V''} \left(d\varphi- \dot{V}' \eta_0 \right)^2\right. \\
 &+ \left. \frac{1-cy_0^2}{12y_0^2} \left(\frac{dw^2}{c_0-c w^2} + (c_0 -c w^2) d\phi^2 \right) \right]. \label{n=asource}
\end{align}  The overall warp factor goes to a constant, $H=\frac{1}{4y_0^2}$.  The one-form $\eta_0$ is given as
\begin{equation}
\eta_0 = d\hat{\chi} + c w d\phi.  
\end{equation} The $(w,\phi)$ describe a two-dimensional surface with curvature $c$.  In the case when $c=-1$, the surface is an $\mathbb{H}_2$ plane which can be replaced with $\mathbb{H}_2/\Gamma$ to obtain a compact Riemann surface with genius $g$.  The $\varphi$ circle shrinks at the location of the source and can be interpreted as an M-theory circle.  The geometry is then an M-theory description of $k$ $D6$-branes wrapped on $AdS_5 \times \Sigma_g$.  Unlike the $\mathcal{N}=2$ case, the $R^3$ normal to the $D6$-branes is non-trivial fibered over the Riemann surface $\Sigma_g$.  

The sources in this case are special and exist only when $n_\pm =a_\pm$.  The orbits of the killing vector $n_+ \frac{\partial}{\partial \phi_+}+ n_- \frac{\partial}{\partial \phi_-}$ shrink at the location of the source, the condition implies that the puncture sits at a point on the $(t^+,t^-)$ plane where the $\mathcal{N}=1$ R-symmetry shrinks.  

One can consider placing more than one of these $D6$-branes on the $(t^+,t^-)$ plane and at the same point on the Riemann surface.  Unlike the $\mathcal{N}=2$ $D6$-branes, these sources will interact with each other due to the $\frac{1}{y} \partial_y \left[ \frac{2}{1-cy^2} e^D\right]$ term in \eqref{n=atoda}.  It is currently unclear what are the effects of this term and how it will fix the location of the sources.  This question need to be explored in order to understand how superpose the $\mathcal{N}=1$ punctures.

\subsection{$\mathcal{N}=1$ Punctures and $M9$-branes}\label{N1M9}

In this section, we discuss the piece-wise punctures when $n_+=-n_-=1$.  The reduction for this case is discussed in section \ref{np=-nmLLM} and equation \eqref{stepeq} reduces to 
\begin{equation}
\frac{1}{r} \partial_r \left(r \partial_r V\right) +\partial_t^2 \left( (1-2ct)^{3/2} V \right) = -4\pi (n_2-n_1) \delta(t-t_0)\delta^2(r) \label{6branesM9}
\end{equation} where $V = - \partial_t^2 \widetilde{D}$.  We analyse the source for when $c=0$.  The behaviour of the source will be same even when $c$ is non-vanishing; the major different is we would need to drop subleading terms to the Laplace operator acting on $V$.  The solution for $V$ is given as 
\begin{equation}
V = \frac{n_2-n_1}{\sqrt{r^2 + (t-t_0)^2}}.  
\end{equation} The metric for the system is 
\begin{align}
ds^2_{11} &= H^{-1/3} \left[ ds^2_{AdS_5} +\frac{1}{3} H ds^2_6 \right] \\
ds^2_6 &= V \left(dt^2 + dr^2 +r^2 d\varphi^2 \right) + \frac{1}{V} \left(d\chi + \omega\right)^2 + 6\left(\frac{dw^2}{1-c_1 w^3} + \frac{4}{9} \frac{1-c_1 w^3}{c_1 w} d\phi^2 \right)
\end{align} with
\begin{equation}
H= \frac{c_1 w}{8}, \qquad \partial_t \omega = -r\partial_r V d\varphi.  
\end{equation}  When $c_1$ vanishes, the metric is degenerate.  When $c_1$ is non-vanishing, it is singular at $w=0$ since the warp factor vanishes.  This singularity can be understood as co-dimension one source in M-theory.  In the region near $w=0$, the metric can be rewritten as
\begin{align}
ds^2_{11} &= H^{-1/3} \left(ds^2_{AdS_5} + \frac{1}{9} d\phi^2 \right)  \\
&+ \frac{1}{3} H^{4/3} \left(  V \left(dt^2 + dr^2 +r^2 d\varphi^2 \right) + \frac{1}{V} \left(d\chi + \omega\right)^2 + 6 dw^2 \right).
\end{align} 
It describes $M5$-branes wrapped on $AdS_5 \times S^1$ \cite{Gueven:1992hh}.  The circle is coordinatized by $\phi$ and is dual to the R-symmetry of the system.  The branes are localized at $w=0$ but smeared along the $(t,r,\varphi,\chi)$ directions, this is indicated by the linear property of the warp factor.  Since $c_1$ is an arbitrary constant, we take it to be a positive parameter, $c^R$, when $w>0$ but a negative parameter, $c^L$, when $w<0$ and write the warp factor as
\begin{equation}
H = \frac{w}{8} \left( c^R  \theta(w) + c^L  \theta(-w) \right)
\end{equation} In this region near $w=0$, warp factor satisfies
\begin{equation}
\Delta_\perp H = \frac{c^R -c^L}{16} \delta(w).  
\end{equation} where $\Delta_\perp$ is the Laplacian along the directions transverse to the $M5$-branes.  The number of $M5$ branes is given by $(c^R-c^L)$ in units fixed by flux quantization.  We refer to these smeared branes as $M9$ branes.  

In the region near the source in \eqref{6branesM9} the $(t,r,\varphi,\chi)$ directions make an $\mathbb{R}^4/\mathbb{Z}_k$ where $k = 2(n_2-n_1)$.  The source correspond to $k$ $D6$-branes in M-theory with $AdS_5 \times S^2$ world volume.  At he location of the branes, the circle dual to the flavor symmetry generated by $\frac{\partial}{\partial \phi_+} - \frac{\partial}{\partial \phi_-}$ shrinks.  The $S^2$ in this case is made from two cups glued together at $w=0$ by the $M9$ branes.  The circle is dual to the $U(1)$ R-symmetry in the field theory.  Regularity conditions in regions where $\phi$ shrinks implies that $c^L= -c^R$.  

In the region near the source, the $M9$-branes should reduce to $D8$-branes.  Generally the linear term in the warp factor can be replaced by a piece-wise linear function in $w$; this corresponds to introducing multiple sets of $M9$-branes in the system.  These $M9$ branes can be added in the $\mathcal{N}=2$ systems described in \ref{N2LLM}, One sees this from the linear term in the warp factor in \eqref{N2LLMwarp}.  So far we have just provided qualitative evidence for these $M9$-branes.  We explore them further for this class solutions and their presences in the $\mathcal{N}=2$ systems on a separate publication.

\section{Summary and Outlook}\label{conclude}

The main goal of this paper is to provide a gravitational description of four-dimensional $\mathcal{N}=1$ SCFTs in the low-energy limit of a stack of $M5$-branes wrapped on a punctured Riemann surface.  We restrict to cases where the surface is embedded in a Calabi-Yau threefold that is a sum two line bundles.  To classify these theories, we look for supersymmetric $AdS_5$ solutions in M-theory where the internal six-dimensional manifold is a four-manifold fibered over a Riemann surface with a $U(1)^2$ structure group.  The reduction of the supergravity equations under such ansatz was initiated in \cite{Bah:2013qya}.  In this paper, we further reduce the system and show that the metric is governed by a single potential $D_0$ which satisfies a generalization of the Monge-Amp\`ere equation \eqref{MAeq}.  The general metric is described in section \ref{Gravitydual}.  

The $U(1)$s correspond to the phases of the two line bundles on the Riemann surface, their generators are denoted as $J^\pm= \frac{\partial}{\partial \phi_\pm}$.  The superconformal R-symmetry of the dual field theory is given as $a_+ J^+ + a_-J^-$ where $a_\pm$ are explicit parameters in the system that are fixed by solutions of $D_0$.  These parameters are normalized as $a_+ + a_-=1$ and can be written as $a_\pm =\frac{1\pm \epsilon}{2}$.  In the dual field theories $\epsilon$ fixes the anomalous dimensions of protected operators and is determined by a-maximization \cite{Intriligator:2003jj}.  The derived system also admits an S-duality that relate different solutions.  We described the action of the duality  in section \ref{Gravitydual}.  

In the special case when the Riemann surface is compact, and has no boundaries, the potential $D_0$ separates into two pieces (these are describe in section \ref{CCSystems}).  The first term is fixed by the Liouville equation and leads to constant curvature surfaces.  The second term in $D_0$ is fixed by a Monge-Amp\`ere equation on a plane normal to the Riemann surface.  Known solutions such as the $Y_{pq}$, GMSW \cite{Gauntlett:2004yd,Gauntlett:2004zh} and $B^3W$ \cite{Bah:2012dg} fit in this class (this was also described in \cite{Bah:2013qya}).  

The ansatz for $D_0$ that yields $B^3W$ solutions can be generalized.  It leads to a family of metrics where the four-dimensional space normal to the Riemann surface is a hyper-surface in $\mathbb{R} \times (\mathbb{R} \times S^1)^{k+1}$.  When $k=2$ we find a family of solutions that are related to $B^3W$ under the S-duality.  When $k>2$, we cannot solve the equation.  We interpret this as the failure of the ansatz to capture the presence of sources localized on a plane normal to the Riemann surface.  In order to fully understand these more general metrics, a B\"acklund transformation on the Monge-Amp\`ere equation is required.  In particular we hope that these systems can provide the gravity duals of six-dimensional $(1,0)$ SCFTs wrapped on Riemann surfaces.  The $(1,0)$ SCFTs were recently classified in \cite{Heckman:2013pva}.  We explore these questions in the future.  

When the Riemann surface has punctures, one needs to provide boundary conditions for the metric.  The spectrum of such boundary conditions appear as different sources for the $D_0$ field that are localized on the Riemann surface.  The punctures can be labelled by their profile along the directions normal to the Riemann surface.  The classification problem of punctures in gravity reduces to studying the set of profiles that lead to regular solutions.  

Near each source, a circle on the $U_+(1)\times U_-(1)$ torus shrinks.  This leads to a natural $(n_+,n_-)$ labelling of the punctures which identifies the shrinking circle as the dual of the killing vector $n_+ J^++ n_- J^-$.  In the region near the puncture, there is also a shrinking circle on the Riemann surface which can be interpreted as a M-theory circle.  We find that the metric in region describes $k$ $D6$-branes. 

In the generic case where $n_+ +n_- =1$, we find $D6$-branes wrapping the overall $AdS_5$ factor and a round $S^2$.  We can write the parameters as $n_+=\frac{p}{p+q}$, $n_- = \frac{q}{p+q}$ where the $p$ and $q$ are co-prime.  The metric in this region preserves eight supercharges in $AdS_5$ and leads to an $SU(k)$ global symmetry in field theory.  For a fixed choice of $(p,q)$ one can superpose these $D6$-branes on the same point on the Riemann surface to obtain a general $\mathcal{N}=2$ puncture that is labelled by a Young Diagram \cite{Gaiotto:2009gz}.  The $(p,q)$ labelling is a generalization of the $\mathcal{Z}_2$ graded punctures that appear in field theory in \cite{Gadde:2013fma,Bah:2013aha,Agarwal:2014rua}.  The elements of the $\mathbb{Z}_2$ is $(1,0)$ and $(0,1)$.

 When $n_\pm =a_\pm$, the sources sit at places where the circle dual to the R-symmetry shrinks.  In this case, the $D6$-branes wrap the $AdS_5$ factor and a Riemann surface $\Sigma_g$.  These punctures locally preserve $\mathcal{N}=1$ supersymmetry.

 When $n_+=-n_-=1$, the shrinking circle is dual to the flavor symmetry $J^+ -J^-$.  In this case, the metric is non-degenerate only when there are co-dimension one sources in the space.  These $M9$-branes appear as smeared $M5$-branes that wrap the $AdS_5$ factor and the $S^1$ dual to the R-symmetry generator $a_+ J^++a_-J^-$.  The $D6$-branes sources wrap the $AdS_5$ factor and an $S^2$ build from two cups that end on the $M9$ branes.  In IIA limit, these $M9$-branes can be interpreted as $D8$-branes.  The $M9$-branes can also be turned on in the generic $(p,q)$ punctures.  The presence of the $M9$-branes reduces the supersymmetry to $\mathcal{N}=1$.  In this paper, we discuss the qualitative features of the $M9$-branes and highlight their necessity in obtaining these $\mathcal{N}=1$ punctures.  We study them on a separate publication.  It should be interesting to understand how these solutions with $M9$-branes are related to the $AdS_7$ solutions with $D8$-branes discussed in \cite{Apruzzi:2013yva}.  
 
 Formally one can take linear combinations of the $D6$-branes profiles to construct a more general $\mathcal{N}=1$ puncture.  Such punctures will preserve the $U_+(1) \times U_-(1)$ symmetry and will contain intersecting $D6$-branes.  In order to understand the physics of these more general objects, we need to describe these intersection regions and find their consistency constraints.  This requires better control of the generalized Monge-Amp\`ere equation that underlie these systems.  
 
 In order to have a complete picture, we need to find interpolating solutions between the puncture geometry and a general $(p,q)$ solution of $B^3W$.  In this paper, we have identified $D_0$ at the end points.  These interpolating solutions can be searched for numerically since we have the general equation and the boundary solutions.  In order to compute central charge and physical quantities, we need to reduce the four form flux of these system and work out the flux quantization.  One needs to start with the expression for the flux as given in \cite{Gauntlett:2004zh} and reduce it through the ansatz in \cite{Bah:2013qya} and in this paper.  This is work in progress.
 
We consider the work in this paper to be a step to systematically study flavor symmetries in AdS/CFT in $\mathcal{N}=1$ backgrounds.  More importantly it would be interesting and useful to be able to read off the quiver gauge theory dual to the gravity solution from the choice of punctures.  Indeed these quivers will include $T_N$ theories \cite{Gaiotto:2009we}.  We also explore these ideas in the future.




\acknowledgments

I would like to thank Vasilis Stylianou for going over the draft.  I would also like to thank Phil Szepietowski, Brian Wecht, Nikolay Bobev, Iosif Bena, Nicholas Warner, Hagen Triendl, Jaewon Song, Ken Intriligator, Prarit Agarwal, John MacGreevy, Yuji Tachikawa, Kazunobu Maruyoshi, Maxime Gabella and Nicholas Halmagyi for useful discussions.  IB is supported in part by the DOE grant DE-SC0011687, ANR grant 08-JCJC-0001-0, and the ERC Starting Grants 240210-String-QCD-BH, and 259133-ObservableString.  I am grateful for the hospitality and work space provided by the UCSD Physics Department.  I am also thankful for the hospitality of KITP and the organizers of the program New Methods in Nonperturbative Quantum Field Theory.

\appendix

\section{Generalized LLM}\label{GenLLM}

Motivated by the $\mathcal{N}=2$ reduction, we can reduce the general $\mathcal{N}=1$ system under the ansatz
\begin{equation}
D_0 = \widetilde{D}(t,x_1,x_2) + F(t^+,t^-) \label{D0texact}
\end{equation} where 
\begin{equation}
t= n_+ t^+ + n_- t^-.
\end{equation} The $n_\pm$ parameters can be restricted to satisfy $n_+ + n_- = 1$ or $n_+ + n_- = 0$.  This ansatz will lead to different embeddings of LLM into the Monge-Amp\`ere system; these can also be obtained by acting with the symmetry \eqref{symmetry} on \eqref{N2D0}.  We also expect a new class of metrics similar to LLM but preserve $\mathcal{N}=1$ supersymmetry.  The systems obtained here will be useful in our discussion of punctures in section \ref{puncD6}.  In this reduction, we will hold the potential $\widetilde{D}$ generic and determine the function $F$.  

The assumption that $\widetilde{D}$ is generic restricts the supersymmetry equation as 
\begin{equation}
 \left(\partial_{x_1}^2 + \partial_{x_2}^2 \right) \widetilde{D} = \left(h_0(t) - h_1(t) \partial_t^2 \widetilde{D} \right) e^{(n_+ + n_-) \partial_t \widetilde{D}}.  \label{GenLLMeq}
\end{equation} The $h$ coefficients are related to $F$ as
\begin{align}
h_1(t) &= -s^{5/2} e^{\left(\partial_+ + \partial_- \right) F} \left(n_+ \partial_- - n_-\partial_+ \right)^2F \\
h_0(t) &= s^{5/2} e^{\left(\partial_+ + \partial_- \right) F}\left[ \partial_+^2 F \partial_-^2 F - \left(\partial_+ \partial_- F\right)^2 \right].
\end{align}  These equations must be analysed independently for the three cases $n_+=-n_-$, $a_\pm=n_\pm$, and generic $n_++n_-=1$.

\subsection{Systems with  $n_++ n_- =1$}\label{N2LLM}

when $n_+ + n_-=1$ the equations can be solved with the coordinates
\begin{equation}
t= n_+ t^+ + n_- t^-, \qquad u= t^+ -t^-, \qquad t_\pm = t \pm n_\mp u.   
\end{equation}  In these coordinates, $s= t+ a_n u$ where $a_n = a_+ n_- -a_- n_+$.  When $h_1=0$, the system can be reduces to Liouville equation; therefore we will not consider such case.  When $h_1$ is non-vanishing, it can be fixed by hand, we keep it for now.   

When $a_n$ is non-vanishing, the solution for $F$ and $h_0$ can be written as
\begin{align}
h_0 &=-h_1 \partial_t \ln \frac{h_1}{(1-2ct)^{3/2}}  \\
e^{\partial_t F} &=2 h_1 \frac{-c_1 (6s)^{3/2} + 6c (1-2ct)^{1/2} s +  (1-2ct)^{3/2}}{3a_n^2 s^{3/2} (1-2ct)^{3/2}}
\end{align}where $c$ and $c_1$ are integration constants.  
In order to write the metric, we make the coordinate transformation and field redefinition
\begin{equation}
e^{D} = \frac{h_1}{1-2ct} e^{\partial_t \widetilde{D}},\qquad s = \frac{1}{6}w^2 (1-2ct), \qquad g(w) = 1+ cw^2 -c_1 w^3 .
\end{equation} We obtain
\begin{align}
ds^2_{11} &=H^{-1/3} \left[ds^2_{AdS_5} + \frac{1}{3} H  ds^2_6 \right] \\
ds^2_6 &= - \left(c+(1-2ct)\partial_t D\right) \left[\frac{dt^2}{1-2ct}+e^{D} \left(dx_1^2 + dx_2^2 \right)\right]\nonumber \\
&  -\frac{3(c_1 w-c)}{H \left(c+(1-2ct)\partial_t D\right)}\left(d\chi + \frac{1}{2}*d D -\frac{1}{2} \frac{c_1 w}{c_1 w- c} d\phi \right)^2 \nonumber \\ 
&+(1-2ct) \left(\frac{dw^2}{g(w)}  + \frac{g(k)}{c_1 w -c} d\phi^2 \right). \label{N2LLMmet}
\end{align} The warp factor is 
\begin{equation}
H = \frac{3}{4(1-2ct)} \left[c_1 w - c \frac{(1-2ct) \partial_tD}{c+(1-2ct)\partial_t D} \right]. \label{N2LLMwarp}
\end{equation} The circles are
\begin{equation}
\chi = \frac{3a_n-2n_-}{3a_n} \phi_+ + \frac{3a_n+2n_+}{3a_n} \phi_-, \qquad \phi = \frac{2}{3a_n} \left(n_-\phi_+-n_+ \phi_-\right).
\end{equation} The supersymmetry equation is
\begin{equation}
 \left(\partial_{x_1}^2 + \partial_{x_2}^2 \right) D+ \partial_t \left[\left(c+ (1-2ct) \partial_t D \right)e^D\right] =0.
\end{equation}  The parameter $c$ can be restricted to take value in $\{-1,0,1\}$ without lost of generality.  The physics of the system is different depending on whether $c_1$ vanishes.  

First we consider the case when $c_1=0$.  The metric is regular and positive definite when $c=-1$.  In this case, we make the coordinate transformation $y^2 = (1+2t)$ to obtain the LLM metric as described above.  The circle coordinates $(\phi, \chi)$ reduce to $(\phi_+, \phi_-)$.  

When $c_1$ is non vanishing, the $\mathcal{N}=2$ supersymmetry is broken to $\mathcal{N}=1$.  It is then interesting to understand the origin of this supersymmetry breaking.  One clue is how $c_1$ appears in the warp factor, it is the coefficient of linear terms in $w$.  This suggest the presences of a codimension one source.  We elaborate on this point later in this paper.  

\subsection{Systems with  $n_\pm = a_\pm$}\label{n=aLLM}

When $a_n=0$, the solution is
\begin{align}
e^{\partial_t F} =h_1 \frac{c_0 (1-2ct)^2 -c (c_1 -u)^2}{ t^{5/2} (1-2ct)}, \qquad h_0 &= -h_1 \partial_t \ln \frac{h_1(1-2c t)}{t^{5/2}}
\end{align} where $c$, $c_0$ and $c_1$ are integration constants.  We make the transformations
\begin{equation}
e^{D} = \frac{h_1}{2t } e^{\partial_t \widetilde{D}}, \qquad u=c_1 +w(1-2ct), \qquad y^2 =2t.  
\end{equation}  The metric becomes
\begin{align}
ds^2_{11} &=H^{-1/3} \left[ds^2_{AdS_5} + \frac{1}{3} H  ds^2_6 \right] \\
ds^2 &=- y\partial_y W \left[ dy^2 +  e^D \left(dx_1^2 + dx_2^2 \right)\right]- \frac{4y^2}{3+ y\partial_y W} \left(d\chi+ \frac{1}{2}*dD +ck d\phi \right)^2 \nonumber \\
&+ (1-c y^2)\left(\frac{dw^2}{c_0 -c w^2} +(c_0 -c w^2 ) d\phi^2 \right).  \label{n=aMet}
\end{align} The warp factors are
\begin{equation}
H = \frac{1}{4y^2} \frac{3+y \partial_y W }{y \partial_y  W},  \qquad e^W = \frac{1-cy^2}{y^3} e^D.  
\end{equation} The equation of motion the potential, $D$, is 
\begin{equation}
\left(\partial_{x_1}^2 + \partial_{x_2}^2 \right) D + \partial_y^2 e^D = \frac{1}{y} \partial_y \left[ \frac{2}{1-cy^2} e^D\right].      \label{an0eq}
\end{equation} The circles are given as
\begin{equation}
\chi = \phi_+ + \phi_-, \qquad \frac{1}{2} \phi = n_- d\phi_+ - n_- d\phi_+.  
\end{equation}

The solutions in this class are inherently $\mathcal{N}=1$ and are govern by a modified $SU(\infty)$ Toda equation.  The parameter $c_0$ can be fixed to $c_0=1$ without lost of generality.  The $(w,\phi)$ directions combine to a Riemann Surface with curvature $c$.  In the case when $c=1$, the surface is a two-sphere, its isometries are dual to a $SU(2)$ flavor symmetry in field theory.  When $D$ is separable in $y$ and $x$, the solutions reduce to the GMSW solutions described in \cite{Gauntlett:2004zh}.  In general there is a larger class of solutions determined by \eqref{an0eq} that describe geometries with localized sources on the Riemann surface.  

\subsection{Systems with $n_+=- n_- =1$}\label{np=-nmLLM}

When $n_+ + n_-=0$, we can fix $n_+=-n_- =1$ without lost of generality and solve the equations in terms of $s$ and $t=t^+-t^-$.  The solution is
\begin{equation}
e^{\partial_s F} = \frac{2}{3} \frac{(1-2ct)^{3/2} -c_1 s^{3/2}}{s^{3/2}}, \qquad h_1(t) = (1-2ct)^{3/2}.  
\end{equation} where $c$ and $c_1$ are integration constants.  The function $h_0$ can be set to zero without lost of generality.  

In order to write the metric, we make the field redefinition and coordinate transformation:
\begin{equation}
V = - \partial_t^2 \widetilde{D}, 	\qquad s= w^2 (1-2ct).
\end{equation} The metric is then
\begin{align}
ds^2_{11} &=H^{-1/3} \left[ds^2_{AdS_5} + \frac{1}{3} H  ds^2_6 \right] \\
ds^2_6 &= V \left[dt^2+ h_1 \left(dx_1^2 + dx_2^2 \right)\right] +V^{-1} \frac{c_1 w}{8 H (1-2ct)} \left(d\chi + \omega -\frac{2c}{c_1 w} d\phi \right)^2 \nonumber \\
&+ 6 (1-2ct)\left(\frac{dw^2}{1-c_1 w^3}  +  \frac{4}{9} \frac{1-c_1 w^3}{c_1 w} d\phi^2 \right).  
\end{align} The equation of motion for $\widetilde{D}$ yields an equation for the potential $V$  and connection $\omega$ given as
\begin{equation}
 \left(\partial_{x_1}^2 + \partial_{x_2}^2 \right) V + \partial_t^2 \left(h_1 V \right) =0, \qquad \partial_t \omega = -r\partial_r V d\varphi.    
\end{equation} The warp factor is
\begin{equation}
H = \frac{1}{16(1-2ct)}\left[2c_1 w - \frac{12 c^2}{(1-2ct) V} \right].
\end{equation}
The circle coordinates are related to the $\phi_\pm$ as
\begin{equation}
\frac{1}{2}\chi = a_- \phi_+ - a_+ \phi_-, \qquad \frac{1}{2}\phi = \phi_+ + \phi_-.  
\end{equation}

The solutions in this class are regular only when the parameter $c_1$ is non-vanishing.  It also appears in the warp factor as the coefficient of a linear term which may be due to a co-dimension one source.

\section{Derivation of system} \label{metricsystem}

The most general supersymmetric $AdS_5$ metric in M-theory that contains two circles fibered over a two-dimensional Riemann surface \cite{Bah:2013qya} is
\begin{equation}\begin{split}
ds^2_{11} &= H^{-1/3}\left[ds^2_{AdS_5} + \frac{1}{3}  \frac{s^3(1-q\partial_q \Gamma)}{\Sigma G}\left(d\psi+\rho \right)^2 + \frac{1}{3} H ds^2_5 \right] \\
ds^2_5 &= \Sigma e^\Lambda \left(dx_1^2 + dx_2^2 \right) + \frac{G}{1-q\partial_q \Gamma} \left[   \frac{\Sigma}{s}\frac{ds^2}{s^2} + \eta_\tau^2+ \left(d\phi+V^I \right)^2 \right].
\end{split}
\end{equation}  We have set the AdS radius, $L$, to one.  We can reintroduce it by multiplying the metric by an overall $L^{4/3}$.  The forms are
\begin{align}
V^I &= V_0 - *d_2 \Gamma \\
\eta_\tau &= (1-q\partial_q\Gamma) \frac{dq}{q} - s\partial_s \Gamma \frac{ds}{s} \\
\rho &=p  d\phi + \frac{1}{2} *d_2 \Lambda- \frac{1}{2} \frac{q \partial_q\Lambda}{1-q\partial_q \Gamma}  \left(d\phi + V^I \right).
\end{align} The one-form, $V_0$, depends only on the Riemann surface coordinates, $x_i$.  The exterior derivative, $d_2$,  is taken along the $x_i$ directions.  The Hodge star operator acts as $*dx_1= -dx_2$.  
The metric functions are

\begin{align} 
H &= \frac{1}{4 s^2} \left[1- \frac{3s^3 \left(1-q\partial_q \Gamma \right)}{G\Sigma} \right] \\
\Sigma &= -\frac{s^3}G \left[(1-q\partial_q \Gamma) s\partial_s \Lambda + s\partial_s \Gamma q\partial_q \Lambda \right]. \label{sigdef}
\end{align}

The left over equations to solve are
\begin{align}
d_2 G_2 =d_2G &=0, \qquad s\partial_s G = q\partial_q G_2 \label{geqs}\\
s^2 q\partial_q \Lambda &= (1-q\partial_q \Gamma) G_2 + s\partial_s \Gamma G \label{qleq} \\
s\partial_s \left(\Sigma e^\Lambda\right) dR_2 &= G_2 d_2 V^I - s^2d_2 * d_2 \Lambda  \label{ssigeq}\\
q\partial_q \left(\Sigma e^\Lambda\right) dR_2 &= G d_2 V^I \label{qsigeq}
\end{align} where $dR_2= dx_1 \wedge dx_2$. 

The System is governed by two potentials $\Lambda$ and $\Gamma$.

\subsection{Democratic description}

In this section, we want to make the system democratic in the the two circles and functions.  These circle can be coordinatized by $\phi_\pm$.  In terms of them, the killing vectors associated to the R-symmetry circle, $\psi$, and flavor $U(1)$, $\phi$, can be written as
\begin{align}
\partial_{\phi} = \frac{1}{2} \left(\partial_{\phi_+} - \partial_{\phi_-} \right), \qquad \partial_\psi =a_+ \partial_{\phi_+} + a_- \partial_{\phi_-}.
\end{align} We choose the new circle to be such that $a_++a_- =1$.  The one form duals to these circles can be written as
\begin{align}
d\psi = d\phi_+ + d\phi_-, \qquad d\phi = 2\left(a_-d\phi_+- a_+ d\phi_-\right)
\end{align} 

Now we also make the field redefinitions
\begin{align}
\Lambda &=D_+ + D_-, \qquad \Gamma = a_+ D_- -a_- D_+ + \log(q).
\end{align} With these transformations, we find
\begin{align}
\eta_{\pm} &= d\phi_{\pm} + \frac{1}{2} *d D_{\pm}\\
d\phi-*d \Gamma &=2\left(a_- \eta_+ - a_+ \eta_- \right) \\
(1-q\partial_q \Gamma) (d\psi + \rho) &=q\partial_q D_+ \eta_- -q\partial_q D_- \eta_+.
\end{align}

The form equations become
\begin{align}
q\partial_q \left(e^{2A}\right) dR_2 &= a_-G~ d*d D_+ -  a_+G ~d*d D_- \\
s\partial_s \left(e^{2A}\right) dR_2 &=\left(a_- G_2-s^2\right) ~ d*d D_+ -  \left(a_+ G_2+ s^2\right) ~ d*d D_-
\end{align} for 
\begin{equation}
e^{2A} = \Sigma e^{D_+ + D_-}.  
\end{equation}
We can collect these equations to
\begin{equation}
 s^2 G ~ d*dD_{\pm} = \left[(a_\pm G_2 \pm  s^2) q\partial_q -a_\pm G s\partial_s \right] \left(e^{2A}\right) dR_2.
\end{equation} The scalar equation becomes
\begin{equation}
\mathcal{J}_+ = \mathcal{J}_- 
\end{equation} for
\begin{equation}
\mathcal{J}_\pm =   \left[(a_\pm G_2 \pm  s^2) q\partial_q -a_\pm G s\partial_s \right] D_\mp.
\end{equation}  This equation is also the integrability condition of the forms equations.  

There is natural coordinate transformation that is begging to be made:
\begin{equation}
- s^2 G \partial_{\pm} =  \left[(a_\pm G_2 \pm  s^2) q\partial_q -a_\pm G s\partial_s \right]
\end{equation} where the new coordinates are $t^{\pm}$.  The derivatives, $\partial_\pm$, are with respect to these coordinates.  

These transformations imply
\begin{align}
\partial_- D_+ &= \partial_+ D_- \\
d*dD_{\pm}  &= -\partial_\pm \left(e^{2A} \right) dR_2.
\end{align} We discuss the function $\Sigma$ after some simplification.

\subsection*{Implications of coords}

It is useful to define the functions 
\begin{equation}
F_\pm =  \frac{ a_\pm G_2 \pm  s^2}{s^2 G}.  
\end{equation}  We have
\begin{equation}
 \partial_\pm = a_\pm\frac{1}{s} \partial_s - F_{\pm} q\partial_q .
\end{equation} From $F_\pm$ we have
\begin{equation}
G \left[ a_-F_+ -a_+ F_-\right]= 1.
\end{equation}  From the definition of the new partials, we can read off the differentials 
\begin{align}
\frac{dq}{q} &= - F_+ s_+ ds_+ -F_- s_- ds_- \\
 sds &= a_+ dt^+ +a_-  dt^-.
\end{align}  From the differentials we find 
\begin{equation}
 s^2= 2 a_+ t^+ +2 a_- t^-.
\end{equation} and
\begin{equation}
\partial_+ F_- = \partial_- F_+.
\end{equation} We can write the inverse transformations as
\begin{align}
q \partial_q &= G \left(a_+ \partial_- - a_-  \partial_+ \right) \\
\frac{1}{s} \partial_s &=  G \left(F_+ \partial_- - F_- \partial_+ \right) 
\end{align}

\subsection{The metric}

Now we are going to write the metric.  The general form is
\begin{equation}
ds^2_{11} = H^{-1/3} \left[ds^2_{AdS_5} +\frac{1}{3} H \left(e^{2A}\left(dx_1^2 + dx_2^2 \right)+ ds^2_2 + ds^2_{T_2} \right)\right]. 
\end{equation} We can write some of the metric functions as
\begin{equation}
\frac{1-q\partial_q \Gamma}{G} = -a_+ \Delta_+ - a_- \Delta_- 
\end{equation} where
\begin{equation}
\Delta_\pm = \left(a_\pm \partial_\mp - a_\mp \partial_\pm \right) D_\mp.
\end{equation} We also have
\begin{align}
\Sigma=s^5 \Sigma_0 &=  s^5\left[ \partial_+ D_+  \partial_- D_- - \partial_+ D_-  \partial_- D_+  \right] \\
&= \frac{s^5}{a_- a_+} \left[ \Delta_+ \Delta_- + \partial_- D_+ \left(a_+ \Delta_+ + a_- \Delta_-\right) \right].
\end{align}

\subsubsection*{The Tile}

We want to write the metric along the $(s,q)$ directions in the $t^\pm$ coordinates.  We write the one form $\eta_\tau$
\begin{align}
\eta_\tau &=  - \partial_{+} \Gamma   dt^+ -  \partial_{-} \Gamma  dt^-.
\end{align}

The metric along the $t^\pm$ becomes
\begin{equation}
\frac{G}{1-q\partial_q \Gamma} \left[   \frac{\Sigma}{s}\frac{ds^2}{s^2} + \eta_\tau^2\right] = g_{ij}  dt^i  dt^j, \qquad \mbox{for} \qquad g_{ij} = -\partial_i D_j.
\end{equation} The $i,j$ indices take $\pm$.

We can write
\begin{align}
 \det(g_{ij}) &=  \Sigma_0  , \qquad e^{2A} = s^5 \Sigma_0 e^{D_+ +D_-}\\
H & =\frac{2c s^2\Sigma_0 +3( a_+ \Delta_+ + a_-\Delta_-)}{8c s^4 \Sigma_0} \\
&= \frac{ s^2 \Delta_+ \Delta_- + \left(a_+ \Delta_+ +a_-\Delta_- \right) \left(3a_+ a_- + s^2 \partial_- D_+ \right) }{4 s^4 a_+ a_- \Sigma_0}.
\end{align}

\subsubsection*{The Torus}

The metric on the Torus can be written as
\begin{align}
ds^2_{T_2} &=  -4 \frac{1}{det(\hat{h}_{ij})} \hat{h}_{ij} \eta_{i} \eta_j \qquad \mbox{with} \qquad \eta_i = d\phi_i + \frac{1}{2} *dD_i 
\end{align} where
\begin{align}
\hat{h}_{++} &= \frac{3 a_-^2 }{s^2}+    \partial_- D_-, \qquad \hat{h}_{+-} = -\frac{3a_+ a_- }{ s^2} -   \partial_- D_+ \\
\hat{h}_{--} &= \frac{3a_+^2 }{ s^2}+   \partial_+ D_+, \qquad \hat{h}_{-+} = -\frac{3a_+ a_- }{s^2} -    \partial_+ D_-.  
\end{align} We also have
\begin{equation}
\Delta_{\pm} = a_{\pm} \hat{h}_{\pm \pm} + a_\mp \hat{h}_{+-}.
\end{equation} It is important to observe that
\begin{equation}
a_+ a_- \det(\hat{h}_{ij}) = \Delta_+ \Delta_- - (a_+ \Delta_+ +a_- \Delta_-) \hat{h}_{+-}.
\end{equation} So we can write
\begin{equation}
H = \frac{1}{4 s^2} \frac{\det(\hat{h}_{ij})}{\det(g_{ij})}
\end{equation}

\subsubsection*{Full Metric}

Now we introduce the potential $D_0$ and write $D_\pm$ as
\begin{equation}
D_\pm = \partial_\pm D_0.
\end{equation} We also redefine $s$ as $s = a_+ t^+ + a_- t^-$ and write the metric coefficients as
\begin{equation}
g_{ij} = -\partial_i \partial_j D_0, \qquad h_{ij} = - \partial_{i} \partial_j \left(D_0 + \frac{3}{2} s \left(\ln s -1 \right) \right).  
\end{equation}  The metric can written as
\begin{equation}
ds^2_{11} = L^{4/3} H^{-1/3} \left[ ds^2_{AdS_5} + \frac{1}{3} H \left(e^{2A} \left(dx_1^2 +dx_2^2\right) +4 h^{ij}\eta_i \eta_j + g_{ij} dt^i  dt^j \right) \right].
\end{equation} The system is determined by a single potential $D_0$ which satisfies
\begin{equation}
\left(\partial_{x_1}^2+\partial_{x_2}^2 \right) D_0 = e^{2A}
\end{equation} The various functions as
\begin{equation}
e^{2A} = s^{5/2} \det(g_{ij}) e^{(\partial_+ + \partial_-)D_0}, \qquad H = \frac{1}{8 s} \frac{\det(h_{ij})}{\det(g_{ij})}.
\end{equation}

\bibliographystyle{utphys}
\bibliography{miracle}

\providecommand{\href}[2]{#2}\begingroup\raggedright\begin{thebibliography}{10}

\bibitem{Witten:1997sc}
E.~Witten, ``{Solutions of four-dimensional field theories via M theory},''
  \href{http://dx.doi.org/10.1016/S0550-3213(97)00416-1}{{\em Nucl.Phys.}
  {\bfseries B500} (1997) 3--42},
\href{http://arxiv.org/abs/hep-th/9703166}{{\ttfamily arXiv:hep-th/9703166
  [hep-th]}}.

\bibitem{Witten:1998jd}
E.~Witten, ``{Branes and the dynamics of QCD},''
\href{http://dx.doi.org/10.1016/S0920-5632(98)00155-8}{{\em
  Nucl.Phys.Proc.Suppl.} {\bfseries 68} (1998) 216--239}.

\bibitem{Hori:1997ab}
K.~Hori, H.~Ooguri, and Y.~Oz, ``{Strong coupling dynamics of four-dimensional
  N=1 gauge theories from M theory five-brane},'' {\em Adv.Theor.Math.Phys.}
  {\bfseries 1} (1998) 1--52,
\href{http://arxiv.org/abs/hep-th/9706082}{{\ttfamily arXiv:hep-th/9706082
  [hep-th]}}.

\bibitem{Giveon:1997sn}
A.~Giveon and O.~Pelc, ``{M theory, type IIA string and 4-D N=1 SUSY SU(N(L)) x
  SU(N(R)) gauge theory},''
  \href{http://dx.doi.org/10.1016/S0550-3213(97)00687-1}{{\em Nucl.Phys.}
  {\bfseries B512} (1998) 103--147},
\href{http://arxiv.org/abs/hep-th/9708168}{{\ttfamily arXiv:hep-th/9708168
  [hep-th]}}.

\bibitem{Gaiotto:2009we}
D.~Gaiotto, ``{N=2 dualities},''
  \href{http://dx.doi.org/10.1007/JHEP08(2012)034}{{\em JHEP} {\bfseries 1208}
  (2012) 034},
\href{http://arxiv.org/abs/0904.2715}{{\ttfamily arXiv:0904.2715 [hep-th]}}.

\bibitem{Gaiotto:2009hg}
D.~Gaiotto, G.~W. Moore, and A.~Neitzke, ``{Wall-crossing, Hitchin Systems, and
  the WKB Approximation},''
\href{http://arxiv.org/abs/0907.3987}{{\ttfamily arXiv:0907.3987 [hep-th]}}.

\bibitem{Tachikawa:2013kta}
Y.~Tachikawa, ``{N=2 supersymmetric dynamics for pedestrians},''
  \href{http://dx.doi.org/10.1007/978-3-319-08822-8}{{\em Lect.Notes Phys.}
  {\bfseries 890} (2013) 2014},
\href{http://arxiv.org/abs/1312.2684}{{\ttfamily arXiv:1312.2684 [hep-th]}}.

\bibitem{Gaiotto:2009gz}
D.~Gaiotto and J.~Maldacena, ``{The Gravity duals of N=2 superconformal field
  theories},''
\href{http://arxiv.org/abs/0904.4466}{{\ttfamily arXiv:0904.4466 [hep-th]}}.

\bibitem{Lin:2004nb}
H.~Lin, O.~Lunin, and J.~M. Maldacena, ``{Bubbling AdS space and 1/2 BPS
  geometries},'' \href{http://dx.doi.org/10.1088/1126-6708/2004/10/025}{{\em
  JHEP} {\bfseries 0410} (2004) 025},
\href{http://arxiv.org/abs/hep-th/0409174}{{\ttfamily arXiv:hep-th/0409174
  [hep-th]}}.

\bibitem{Maldacena:1997re}
J.~M. Maldacena, ``{The Large N limit of superconformal field theories and
  supergravity},'' {\em Adv.Theor.Math.Phys.} {\bfseries 2} (1998) 231--252,
\href{http://arxiv.org/abs/hep-th/9711200}{{\ttfamily arXiv:hep-th/9711200
  [hep-th]}}.

\bibitem{Bah:2011je}
I.~Bah and B.~Wecht, ``{New N=1 Superconformal Field Theories In Four
  Dimensions},'' \href{http://dx.doi.org/10.1007/JHEP07(2013)107}{{\em JHEP}
  {\bfseries 1307} (2013) 107},
\href{http://arxiv.org/abs/1111.3402}{{\ttfamily arXiv:1111.3402 [hep-th]}}.

\bibitem{Bah:2012dg}
I.~Bah, C.~Beem, N.~Bobev, and B.~Wecht, ``{Four-Dimensional SCFTs from
  M5-Branes},'' \href{http://dx.doi.org/10.1007/JHEP06(2012)005}{{\em JHEP}
  {\bfseries 1206} (2012) 005},
\href{http://arxiv.org/abs/1203.0303}{{\ttfamily arXiv:1203.0303 [hep-th]}}.

\bibitem{Bah:2013aha}
I.~Bah and N.~Bobev, ``{Linear quivers and $ \mathcal{N} $ = 1 SCFTs from
  M5-branes},'' \href{http://dx.doi.org/10.1007/JHEP08(2014)121}{{\em JHEP}
  {\bfseries 1408} (2014) 121},
\href{http://arxiv.org/abs/1307.7104}{{\ttfamily arXiv:1307.7104}}.

\bibitem{Benini:2009mz}
F.~Benini, Y.~Tachikawa, and B.~Wecht, ``{Sicilian gauge theories and N=1
  dualities},'' \href{http://dx.doi.org/10.1007/JHEP01(2010)088}{{\em JHEP}
  {\bfseries 1001} (2010) 088},
\href{http://arxiv.org/abs/0909.1327}{{\ttfamily arXiv:0909.1327 [hep-th]}}.

\bibitem{Maruyoshi:2009uk}
K.~Maruyoshi, M.~Taki, S.~Terashima, and F.~Yagi, ``{New Seiberg Dualities from
  N=2 Dualities},'' \href{http://dx.doi.org/10.1088/1126-6708/2009/09/086}{{\em
  JHEP} {\bfseries 0909} (2009) 086},
\href{http://arxiv.org/abs/0907.2625}{{\ttfamily arXiv:0907.2625 [hep-th]}}.

\bibitem{Beem:2012yn}
C.~Beem and A.~Gadde, ``{The superconformal index of N=1 class S fixed
  points},''
\href{http://arxiv.org/abs/1212.1467}{{\ttfamily arXiv:1212.1467 [hep-th]}}.

\bibitem{Gadde:2013fma}
A.~Gadde, K.~Maruyoshi, Y.~Tachikawa, and W.~Yan, ``{New N=1 Dualities},''
  \href{http://dx.doi.org/10.1007/JHEP06(2013)056}{{\em JHEP} {\bfseries 1306}
  (2013) 056},
\href{http://arxiv.org/abs/1303.0836}{{\ttfamily arXiv:1303.0836 [hep-th]}}.

\bibitem{Xie:2013gma}
D.~Xie, ``{M5 brane and four dimensional N = 1 theories I},''
  \href{http://dx.doi.org/10.1007/JHEP04(2014)154}{{\em JHEP} {\bfseries 1404}
  (2014) 154},
\href{http://arxiv.org/abs/1307.5877}{{\ttfamily arXiv:1307.5877 [hep-th]}}.

\bibitem{Xie:2013rsa}
D.~Xie and K.~Yonekura, ``{Generalized Hitchin system, Spectral curve and
  $\mathcal{N} = 1$ dynamics},''
  \href{http://dx.doi.org/10.1007/JHEP01(2014)001}{{\em JHEP} {\bfseries 1401}
  (2014) 001},
\href{http://arxiv.org/abs/1310.0467}{{\ttfamily arXiv:1310.0467 [hep-th]}}.

\bibitem{Xie:2014yya}
D.~Xie, ``{N=1 Curve},''
\href{http://arxiv.org/abs/1409.8306}{{\ttfamily arXiv:1409.8306 [hep-th]}}.

\bibitem{McGrane:2014pma}
J.~McGrane and B.~Wecht, ``{Theories of Class S and New N=1 SCFTs},''
\href{http://arxiv.org/abs/1409.7668}{{\ttfamily arXiv:1409.7668 [hep-th]}}.

\bibitem{Agarwal:2013uga}
P.~Agarwal and J.~Song, ``{New N=1 Dualities from M5-branes and
  Outer-automorphism Twists},''
  \href{http://dx.doi.org/10.1007/JHEP03(2014)133}{{\em JHEP} {\bfseries 1403}
  (2014) 133},
\href{http://arxiv.org/abs/1311.2945}{{\ttfamily arXiv:1311.2945 [hep-th]}}.

\bibitem{Agarwal:2014rua}
P.~Agarwal, I.~Bah, K.~Maruyoshi, and J.~Song, ``{Quiver Tails and N=1 SCFTs
  from M5-branes},''
\href{http://arxiv.org/abs/1409.1908}{{\ttfamily arXiv:1409.1908 [hep-th]}}.

\bibitem{Bonelli:2013pva}
G.~Bonelli, S.~Giacomelli, K.~Maruyoshi, and A.~Tanzini, ``{N=1 Geometries via
  M-theory},'' \href{http://dx.doi.org/10.1007/JHEP10(2013)227}{{\em JHEP}
  {\bfseries 1310} (2013) 227},
\href{http://arxiv.org/abs/1307.7703}{{\ttfamily arXiv:1307.7703}}.

\bibitem{Giacomelli:2014rna}
S.~Giacomelli, ``{Four dimensional superconformal theories from M5 branes},''
\href{http://arxiv.org/abs/1409.3077}{{\ttfamily arXiv:1409.3077 [hep-th]}}.

\bibitem{Bah:2011vv}
I.~Bah, C.~Beem, N.~Bobev, and B.~Wecht, ``{AdS/CFT Dual Pairs from M5-Branes
  on Riemann Surfaces},''
  \href{http://dx.doi.org/10.1103/PhysRevD.85.121901}{{\em Phys.Rev.}
  {\bfseries D85} (2012) 121901},
\href{http://arxiv.org/abs/1112.5487}{{\ttfamily arXiv:1112.5487 [hep-th]}}.

\bibitem{Maldacena:2000mw}
J.~M. Maldacena and C.~Nunez, ``{Supergravity description of field theories on
  curved manifolds and a no go theorem},''
  \href{http://dx.doi.org/10.1142/S0217751X01003937}{{\em Int.J.Mod.Phys.}
  {\bfseries A16} (2001) 822--855},
\href{http://arxiv.org/abs/hep-th/0007018}{{\ttfamily arXiv:hep-th/0007018
  [hep-th]}}.

\bibitem{Bah:2013wda}
I.~Bah, M.~Gabella, and N.~Halmagyi, ``{Punctures from probe M5-branes and $
  \mathcal{N} $ = 1 superconformal field theories},''
  \href{http://dx.doi.org/10.1007/JHEP07(2014)131}{{\em JHEP} {\bfseries 1407}
  (2014) 131},
\href{http://arxiv.org/abs/1312.6687}{{\ttfamily arXiv:1312.6687 [hep-th]}}.

\bibitem{Witten:1988ze}
E.~Witten, ``{Topological Quantum Field Theory},''
\href{http://dx.doi.org/10.1007/BF01223371}{{\em Commun.Math.Phys.} {\bfseries
  117} (1988) 353}.

\bibitem{Bershadsky:1995qy}
M.~Bershadsky, C.~Vafa, and V.~Sadov, ``{D-branes and topological field
  theories},'' \href{http://dx.doi.org/10.1016/0550-3213(96)00026-0}{{\em
  Nucl.Phys.} {\bfseries B463} (1996) 420--434},
\href{http://arxiv.org/abs/hep-th/9511222}{{\ttfamily arXiv:hep-th/9511222
  [hep-th]}}.

\bibitem{Intriligator:2003jj}
K.~A. Intriligator and B.~Wecht, ``{The Exact superconformal R symmetry
  maximizes a},'' \href{http://dx.doi.org/10.1016/S0550-3213(03)00459-0}{{\em
  Nucl.Phys.} {\bfseries B667} (2003) 183--200},
\href{http://arxiv.org/abs/hep-th/0304128}{{\ttfamily arXiv:hep-th/0304128
  [hep-th]}}.

\bibitem{Bah:2013qya}
I.~Bah, ``{Quarter-BPS $AdS_{5}$ solutions in M-theory with a $T^{2}$ bundle
  over a Riemann surface},''
  \href{http://dx.doi.org/10.1007/JHEP08(2013)137}{{\em JHEP} {\bfseries 1308}
  (2013) 137},
\href{http://arxiv.org/abs/1304.4954}{{\ttfamily arXiv:1304.4954 [hep-th]}}.

\bibitem{Gauntlett:2004zh}
J.~P. Gauntlett, D.~Martelli, J.~Sparks, and D.~Waldram, ``{Supersymmetric
  AdS(5) solutions of M theory},''
  \href{http://dx.doi.org/10.1088/0264-9381/21/18/005}{{\em Class.Quant.Grav.}
  {\bfseries 21} (2004) 4335--4366},
\href{http://arxiv.org/abs/hep-th/0402153}{{\ttfamily arXiv:hep-th/0402153
  [hep-th]}}.

\bibitem{Lunin:2008tf}
O.~Lunin, ``{Brane webs and 1/4-BPS geometries},''
  \href{http://dx.doi.org/10.1088/1126-6708/2008/09/028}{{\em JHEP} {\bfseries
  0809} (2008) 028},
\href{http://arxiv.org/abs/0802.0735}{{\ttfamily arXiv:0802.0735 [hep-th]}}.

\bibitem{Ward:1990qt}
R.~Ward, ``{Einstein-Weyl spaces and SU(infinity) Toda fields},''
\href{http://dx.doi.org/10.1088/0264-9381/7/4/003}{{\em Class.Quant.Grav.}
  {\bfseries 7} (1990) L95--L98}.

\bibitem{Gueven:1992hh}
R.~Gueven, ``{Black p-brane solutions of D = 11 supergravity theory},''
\href{http://dx.doi.org/10.1016/0370-2693(92)90540-K}{{\em Phys.Lett.}
  {\bfseries B276} (1992) 49--55}.

\bibitem{Gauntlett:2004yd}
J.~P. Gauntlett, D.~Martelli, J.~Sparks, and D.~Waldram, ``{Sasaki-Einstein
  metrics on S**2 x S**3},'' {\em Adv.Theor.Math.Phys.} {\bfseries 8} (2004)
  711--734,
\href{http://arxiv.org/abs/hep-th/0403002}{{\ttfamily arXiv:hep-th/0403002
  [hep-th]}}.

\bibitem{Heckman:2013pva}
J.~J. Heckman, D.~R. Morrison, and C.~Vafa, ``{On the Classification of 6D
  SCFTs and Generalized ADE Orbifolds},''
  \href{http://dx.doi.org/10.1007/JHEP05(2014)028}{{\em JHEP} {\bfseries 1405}
  (2014) 028},
\href{http://arxiv.org/abs/1312.5746}{{\ttfamily arXiv:1312.5746 [hep-th]}}.

\bibitem{Apruzzi:2013yva}
F.~Apruzzi, M.~Fazzi, D.~Rosa, and A.~Tomasiello, ``{All $AdS_7$ solutions of
  type II supergravity},''
  \href{http://dx.doi.org/10.1007/JHEP04(2014)064}{{\em JHEP} {\bfseries 1404}
  (2014) 064},
\href{http://arxiv.org/abs/1309.2949}{{\ttfamily arXiv:1309.2949 [hep-th]}}.

\end{thebibliography}\endgroup

\end{document}